# Fractal-based modeling and spatial analysis of urban form and growth: a case study of Shenzhen in China

Xiaoming Man, Yanguang Chen

(Department of Geography, College of Urban and Environmental Sciences, Peking University, Beijing 100871, P.R. China. E-mail: xiaoming1231@pku.edu.cn; chenyg@pku.edu.cn)

**Abstract:** Fractal dimension curves of urban growth can be modeled with sigmoid functions, including logistic function and quadratic logistic function. Different types of logistic functions indicate different spatial dynamics. The fractal dimension curves of urban growth in western countries follows the common logistic function, while these curves of cities in northern China follows quadratic logistic function. Now we want to know whether all Chinese cities follow the same rules of urban evolution. This paper is devoted to exploring the fractals and fractal dimension properties of the city of Shenzhen in southern China. The urban region is divided into four subareas, ArcGIS technology, box-counting method is adopted to extract spatial datasets, and the least squares regression is employed to estimate fractal parameters. The results show that: (1) The urban form of Shenzhen city has clear fractal structure, but fractal dimension values of different subareas are different; (2) The fractal dimension growth curves of all the four study areas can only be modeled by the common logistic function, and the goodness of fit increases over time; (3) The peak of urban growth in Shenzhen had passed before 1986, the fractal dimension growth is approaching its maximum capacity. Conclusions can be reached that the urban form of Shenzhen bears characteristics of multifractals, the fractal structure has been becoming better gradually through self-organization, but its land resources are reaching the limits of growth. The fractal dimension curves of Shenzhen's urban growth are similar to those of European and American cities, but differ from those of the cities in northern China. This suggests that there is subtle different dynamic mechanisms of city development between northern and southern China.

# 1. Introduction

A study on cities begins from description and ends at understanding. To describe an urban phenomenon, we have to find its characteristic scales. Traditional mathematical methods and quantitative analysis are based on typical scale, which is often termed *characteristic length* (Hao, 1986; Liu and Liu, 1993; Takayasu, 1990). Unfortunately, spatial patterns of cities has no characteristic scale and cannot be effectively described by conventional measure such as length and area. In this scale, the concept of characteristic scales should be substituted with scaling ideas. Fractal geometry provides a powerful mathematical tools for scaling analysis of urban form and growth. From the remote sensing images, urban form resembles ink splashes, usually presents a highly irregularity and self-similarity at several different scales (Frankhauser, 2004; Benguigui *et al*., 2004). It implies that it does not obey Gaussian law, traditional measures and mathematics models cannot effectively describe it (Salat, 2017; Chen and Huang, 2019). Fractal geometry provides a proper quantitative approach in this aspect (Frankhauser, 1998; Chen *et al*., 2017). The fractal dimension, especially multifractal parameter spectrums, can be utilized to characterize spatial heterogeneity, explore the spatial complexity (Jevric and Romanovich, 2016). Ever since Mandelbrot (1983) developed fractal geometry, the theory has been applied to geographical research for nearly forty years. Urban geography is one of the biggest beneficiaries from fractal ideas (Dauphiné, 2013). Since the 1980's, some pioneering studies about the urban form and growth based on fractal geometry have been published (such as Arlinghaus, 1985; Batty *et al*, 1989; Batty and Longley, 1986; Batty and Longley, 1987a; Batty and Longley,1987b; Batty and Longley,1988; Batty and Xie, 1996; Batty and Kim, 1992; Benguigui *et al*., 2000; Benguigui *et al*, 2001a; Benguigui *et al*, 2001b; Benguigui *et al*, 2004; Encarnação *et al*., 2012; Shen, 2002; White and Engelen, 1993). The research results are once summarized by Batty and Longley (1994) and Frankhauser (1994). Recent years, new progress of studies on fractal cities have been made, and many interesting results were reported in literature (Boeing, 2018; Chen, 2012, 2018, 2020; Chen and Huang, 2019; Lagarias and Prastacos, 2018; Leyton-Pavez *et al*, 2017; Li *et al*, 2013; Ma *et al*, 2020; Man *et al*, 2019; Purevtseren *et al*, 2018; Rastogi and Jain, 2018; ShreevastavaRao and McGrath, 2019; Song and Yu, 2019; Tucek and Janoska, 2013; Versini *et al*, 2020).

Fractal parameters and the mathematical models based on fractal parameters of urban form are keys to understanding the rules of urban evolution in different time and space. Fractal dimension nowadays has been regarded as a validity indicator for assessing the space filling extent, spatial complexity and spatial homogeneity of urban land use patterns. To make spatial analysis of urban form, we need to compare fractal dimension values of different urban regions or different cities. To make dynamic analysis, we have to compare fractal dimension values of a city at different times. A time series of fractal dimension values of a city forms a fractal dimension growth curves. A discovery is that the fractal dimension curves of urban growth takes on squash effect and can be modeled with sigmoid functions (Chen, 2018). The fractal dimension growth curves of urban form in Europe and America satisfy the common logistic function, while those of northern Chinese cities like Beijing meet the quadratic logistic function (Chen, 2012; Chen and Huang, 2019). Different types of logistic functions indicate different spatial dynamics.

Now we want to know whether all Chinese cities follow the same rules of development. How about the cities in southern China? This paper is devoted to explore fractal dimension growth curves of the city of Shenzhen. Shenzhen can be regarded as a shock city in southern China, which became highly booming in the short term after China's reform and opening up. Taking Shenzhen as an example of southern China, we can discuss another type of fractal cities in Mainland China, which differ from many cities in other places of China except for southeast coastal area of China. We first divide Shenzhen city into four study areas, and then use the box-counting method and least squares regression for extracting spatial data and estimating the fractal parameters. Then, we utilize sigmoid functions to model fractal dimension curves of urban growth in Shenzhen region. Through this study, we can not only reveal the North-South differences of city development in China, but also reflect the similarities and differences of urban evolution between China and the West.

## 2. Methods

### 2.1. Box-counting method

Owing to scale-free properties of urban form, the conventional measures should be replaced by fractal parameters. The box-counting method in this paper is employed for estimating the fractal dimension of urban form of four study regions in Shenzhen from 1986 to 2017. It has become a method widely applied by many researchers (such as Mandelbrot, 1967; Batty and Longley, 1994; Benguigui *et al.*, 2000; Shen, 2002; Encarnação *et al.*, 2012; Chen and Wang, 2013; Ni *et al.*, 2017) to measure the fractal dimension in 2-dimensional images. Its basic procedure in general is to recursively superimpose a series of regular grids of declining box sizes over a target object, and then record the object count in each successive box, where the count records how many of the boxes are occupied by the target object. According to Benguigui *et al.* (2000) reported, in a 2-dimensional space, the object is covered by a grid made of squares of size $\varepsilon$, the number $N(\varepsilon)$ of squares in which a part of the object appears is counted. Then changing the side length of boxes, $\varepsilon$, leads to change of nonempty boxes number, $N(\varepsilon)$. if the object turns out to be fractal, then

$$N(\varepsilon) = N_1 (1/\varepsilon)^D \tag{1}$$

where $N_1$ is the proportionality coefficient, $D$ is the urban form fractal dimension value, the logarithmic form is

$$\ln N(\varepsilon) = \ln N_1 + D \ln(1/\varepsilon) \tag{2}$$

Thus, a logarithmic plot of $\ln N(\varepsilon)$ versus $\ln(1/\varepsilon)$ yields a straight line with a slope equal to $D$. This paper, the first value $\varepsilon$ is the half size of the box, the next value is equal to $\varepsilon/4$. The *i*th value is $\varepsilon/2^i$, the highest value of index $i$ is 9. The tools of create fishnet, spatial adjustment, and overlay in ArcMap10.2 were utilized for implementation the segmentation box, rotation box, and obtaining the *i*th value of $N(\varepsilon)$.

## 2.2. Fractal dimension growth curve and power law

However, it is not enough only to compare fractal dimension values from different time periods or regions. The growth characteristics and trends of a city can be interpreted and predicted to a great extent must rely on developing mathematical models. Thus, this paper model fractal dimension values of four study areas in Shenzhen by using the logistic function modeling. According to Chen (2018)'s elaboration, the logistic function of fractal dimension evolution can be expressed as

$$D(t) = \frac{D_{\max}}{1 + (D_{\max}/D_0)e^{-kt}} \quad \text{or} \quad D(t) = \frac{D_{\max}}{1 + Ae^{-k(n-n_0)}} \tag{3}$$

where $t$ is time order (0, 1…), and $n$ to year, $n_0$ to the initial year, $D$(t) or $D$(n) denotes the fractal dimension in the $t$th time or the year of $n$, $D_0$ is the fractal dimension in the initial year, $D_{max} \leqslant 2$ indicates the maximum of the fractal dimension, A refers to a parameter, $k$ is the original growth rate of fractal dimension. The parameter and variable relationships are as follows

$$D(t) = D(n), A = \frac{D_{max}}{D_0} - 1, t = n - n_0 \tag{4}$$

Equation (3) can be made a logarithmic transform, the result is

$$\ln(\frac{D_{max}}{D_0} - 1) = \ln(\frac{D_{max}}{D_0}) - kt = \ln A - kt \tag{5}$$

Equation (5) is concerted to a log-linear equation, and the values of $A$ and $K$ is simply to be estimated by the linear regression analysis if the parameter of $D_{max}$ value is known. Here, the goodness-of-fit search (GOFS) parameter estimation method was selected for estimating the parameter, which has been introduced particularly by Chen (2018).

In addition, for a city system, the relationship between two measure elements of representing city, such as population, area and GDP, usually satisfies the statistical power law. Urban is a typical complex system (Batty, 2008; Batty, 2009). A stable urban form and growth is the result of long-term continues interaction of various factors, those associations can be capture by the power function (Keuschnigg *et al.*, 2019). We thus capture the driving factors of urban form evolution and growth using the power law function, which is expressed as

$$D_t = kX_t^{\beta} \tag{6}$$

where $X_t$ and $D_t$ represent, in a given $t$ year, fractal dimension value and the total number of driving factor, respectively, $k$ and $\beta$ are constants to fractal dimension. The linearized model of equation (6) is

$$\ln D_t = \ln k + \beta \ln X_t \tag{7}$$

in which where ln$k$ and $\beta$ are constants terms to be estimated.

## 2.3. Study area and datasets

Shenzhen (22°27′-22°52′N, 113°46′-114°37′E), as China's first special economic zone, lies along the cost of the South China Sea and adjacent to Hong Kong. Before China's reform and opening-up policy in late 1978, it was just a sleepy border town of some 30,000 inhabitants that served as a custom stop into mainland China from Hong Kong. Now Shenzhen has become an international metropolis with the total population of 13.0366 million (SSB, 2019). Mean annual temperature is around 22.4 ℃ and annual rainfall is about1948 mm (Li *et al*, 2005). Narrow and long is the shape characteristics of the administrative region of Shenzhen, east–west span is over 49 km, while north–south span is only about 7km (Ng, 2003). In the southeastern part of Shenzhen is hilly topography, in the northwestern part is relatively low. This paper, four boxes were drawn as the study areas (**Fig. 1**). The first box area is the entire region of Shenzhen, which almost covers the whole built-up patch of Shenzhen. The second box area is a major center region, which mainly includes Futian District, Luohu District, and Nanshan District. The third and fourth box areas are the northwest part and northeast part. The reason of including them is because the built areas in these two areas visually seem to have the expansion and development trends. It is quite useful to fully assess the spatial-temporal evolution characteristics of the urban form and growth.

The built-up areas data (**Fig. 1**) is extracted from Landsat TM 4, 5, and OLI 8 images with 30m resolution for twelve years: 1986, 1989,1992, 1995, 1998, 2001, 2003, 2006, 2010, 2013, 2015, and 2017. They were all collected from USGS Earth Explorer website (http://earthexplorer.usgs.gov/), with less than 10% cloud cover. In generally, the methods of object-oriented supervised classification and visual interpretation post Classification are employed for extraction the built-up areas data, and its process can be divided into three parts. Firstly, employ the toolbox of example-based feature extracting workflow in ENVI 5.3 software to generate twelve periods land use and land cover (LULC) classification maps, here we roughly divided land types into four categories that are water body, vegetation, bare, and built-up areas, and chose the support vector machine (SVM) classification method. Secondly, select the class of urban area form each resulting vector data in ArcMap10.2 software. Finally, and most crucial, visual interpretation each 12 periods of built-up area maps. It includes removing the noisy patches by setting a threshold value, adding, removing and modifying the misclassification of the built-up areas by using the editor tool and other toolbox in ArcMap 10.2, Google Earth and multi-period false color composite of remote sensing images.

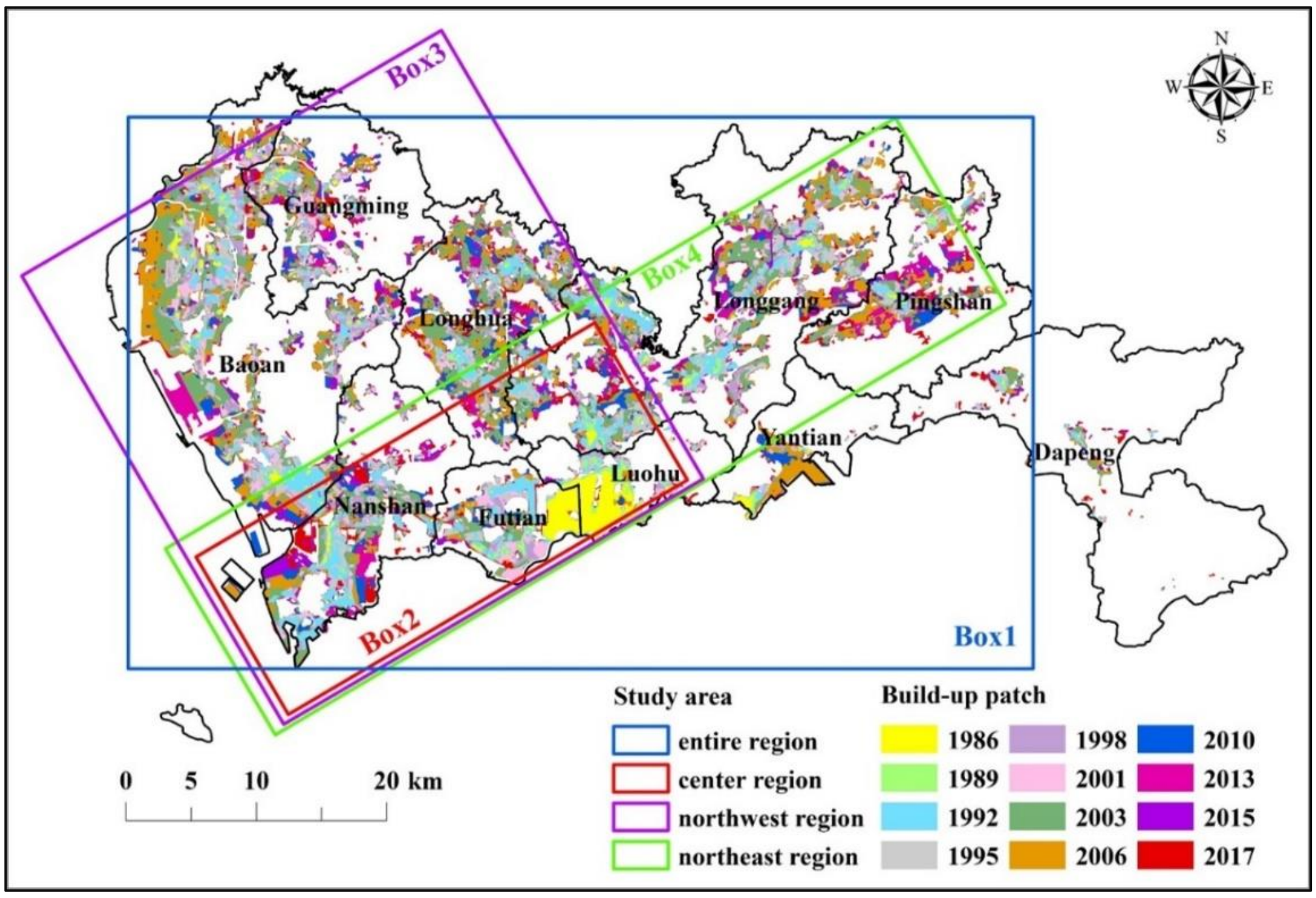

**Figure 1 The built-up area map in Shenzhen city from 1986 to 2017**

**Note:** The built-up areas were extracted from Landsat TM 4, 5, and OLI 8 images with 30m resolution downloading from USGS Earth Explorer website (http://earthexplorer.usgs.gov/). Four boxes are study areas, representing entire region, major business center region, northwest region and northeast region of Shenzhen, respectively.

# 3. Results and analyses

## 3.1. Fractal dimension analysis of urban form

The double logarithm linear regression based on the least squares method can be employed to estimate fractal parameters. Fractal modeling involves two types of parameters. One is inferential parameters, and the other is descriptive parameters. The former includes fractal dimension and proportionality coefficient, and the latter includes goodness of fit and standard error (Chen, 2020;

Shelberg *et al*, 1982). Fitting results based on box-counting method show that, from 1986 to 2017, four study areas all exhibit a strong linear association between $\ln N(\varepsilon)$ and $\ln N(1/\varepsilon)$, with the fractal dimension values between 1 and 2, and the goodness of fit $R^2$ are all above 0.98 (see Appendices). It is a powerful proof indicates that the urban from in Shenzhen, both its entire and part regions, are indeed statistically fractals during 1986-2017. The detailed result of fractal dimension estimates of urban form of four regions in Shenzhen from 1986 to 2017 is presented in **Tab.1**. They are all increased over time gradually, but the entire region is slightly smaller than other three subregions- center region, northwest region and northeast region.

In fact, fractal dimension *D* now has become a validity index of assessing the space filling extent, spatial complexity, and spatial homogeneity or compactness of urban land (Chen and Huang, 2018). The larger value signifies urban sprawl, urban spatial structure becomes more complicated and homogeneous (Islam and Metternicht, 2003). In 2-dimension digital maps, $D = 2$ indicates that one has a homogeneous spatial distribution of the object in the plane, the other extreme limit is $D = 0$, suggests a high local concentration (Benguigui *et al*., 2004). It is obvious from above results, for the past 40 years, the urban form in entire Shenzhen has been in a state of continuing growth and expansion in space. Meanwhile, its spatial structure has been becoming more and more complex and homogeneous over time. If considering the land type distribution in the real world, it is inevitable that there exist other land types such ecological land and water body, in four study areas. They may actually belong to the land types that are protected by the authorities and not allowed to be available for urban construction in Shenzhen. In 2017, in particular, the fractal dimension values of four regions- the entire, central part, northwest part, and northeast part, reached their highest value, which are 1.7604, 1.8467, 1.8189, and 1.8294, respectively (**Tab.1**). Not least because these values are close to 2, but there are probably no more other land types allowed to be available for urban development within each study areas. We thus may speculate that Shenzhen will probably encounter the situation of urban land approaching saturation in the near future.

In addition, we can obtain other further information by calculating the data in **Tab.1**. The average speed of urban space filling degree in the four study areas can be calculated using $(D_{2017} - D_{1986})/\triangle t_{(2017-1986)}$. The calculation result values are 0.0202, 0.0133, 0.0189, and 0.0177, respectively. In terms of area, entire region is the largest region among four study regions, and its value of average speed of urban space filling degree is also the largest. It suggests that Shenzhen experienced large number of other land types to urban land in this period, and the fastest growth and expansion of urban land occurred in the northwest part, followed by the northeast part and central part. Moreover, as shown in **Fig.2**, it is clear that Shenzhen city took place two peaks of urban sprawl during 1986-2017, which are in 1992 and in 2003, but the extent of urban expansion in 1992 is significantly higher in 2003. In 1992, northwest region and northeast region of Shenzhen had the same urban land-use development level, by contract, central region is a slower. But in 2003, These three subregions have basically the same urban land-use development level.

**Table1 Fractal dimension estimates of urban form of four regions in Shenzhen, 1986-2017**

| Year | Entire region | | | Center region | | | Northwest region | | | Northeast region | | |
|---|---|---|---|---|---|---|---|---|---|---|---|---|
| | $D_f$ | $R^2$ | $\sigma$ | $D_f$ | $R^2$ | $\sigma$ | $D_f$ | $R^2$ | $\sigma$ | $D_f$ | $R^2$ | $\sigma$ |
| **1986** | 1.1352 | 0.9886 | 0.0460 | 1.4353 | 0.9921 | 0.0483 | 1.2320 | 0.9914 | 0.0434 | 1.2808 | 0.9912 | 0.0455 |
| **1989** | 1.2101 | 0.9908 | 0.0442 | 1.4730 | 0.9950 | 0.0395 | 1.3037 | 0.9947 | 0.0358 | 1.3379 | 0.9934 | 0.0411 |
| **1992** | 1.4208 | 0.9971 | 0.0289 | 1.6073 | 0.9991 | 0.0186 | 1.4999 | 0.9984 | 0.0228 | 1.5333 | 0.9982 | 0.0245 |
| **1995** | 1.5008 | 0.9983 | 0.0235 | 1.6668 | 0.9995 | 0.0136 | 1.5739 | 0.9991 | 0.0175 | 1.5999 | 0.9985 | 0.0233 |
| **1998** | 1.5454 | 0.9989 | 0.0193 | 1.6866 | 0.9996 | 0.0135 | 1.6174 | 0.9994 | 0.0154 | 1.6311 | 0.9987 | 0.0225 |
| **2001** | 1.5856 | 0.9992 | 0.0171 | 1.7145 | 0.9997 | 0.0121 | 1.6568 | 0.9994 | 0.0149 | 1.6662 | 0.9990 | 0.0200 |
| **2003** | 1.6619 | 0.9995 | 0.0144 | 1.7636 | 0.9998 | 0.0103 | 1.7315 | 0.9997 | 0.0122 | 1.7304 | 0.9993 | 0.0176 |
| **2006** | 1.7050 | 0.9996 | 0.0125 | 1.7884 | 0.9998 | 0.0096 | 1.7691 | 0.9997 | 0.0106 | 1.7705 | 0.9995 | 0.0151 |
| **2010** | 1.7239 | 0.9997 | 0.0117 | 1.8053 | 0.9998 | 0.0106 | 1.7838 | 0.9997 | 0.0110 | 1.7898 | 0.9995 | 0.0145 |
| **2013** | 1.7401 | 0.9998 | 0.0104 | 1.8174 | 0.9998 | 0.0097 | 1.7871 | 0.9998 | 0.0103 | 1.8063 | 0.9996 | 0.0131 |
| **2015** | 1.7489 | 0.9998 | 0.0098 | 1.8309 | 0.9999 | 0.0082 | 1.8076 | 0.9998 | 0.0095 | 1.8166 | 0.9997 | 0.0124 |
| **2017** | 1.7604 | 0.9998 | 0.0089 | 1.8467 | 0.9999 | 0.0075 | 1.8189 | 0.9998 | 0.0088 | 1.8294 | 0.9997 | 0.0115 |

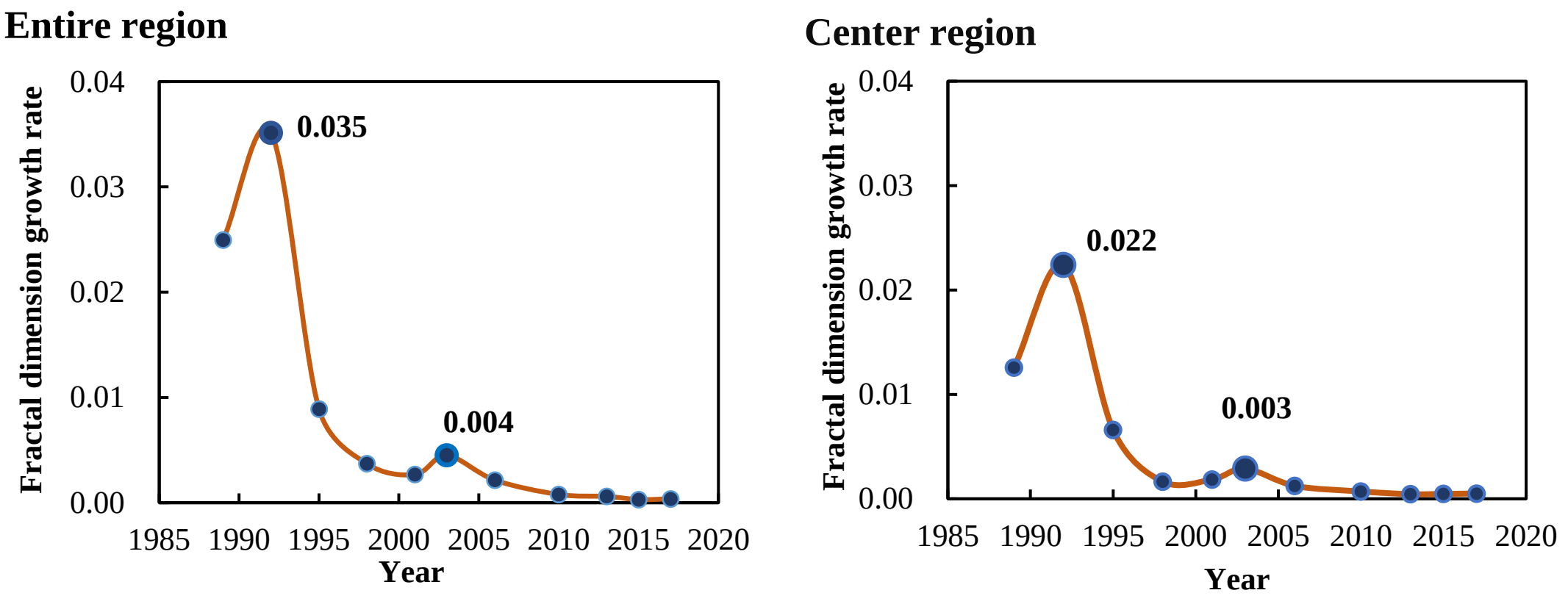

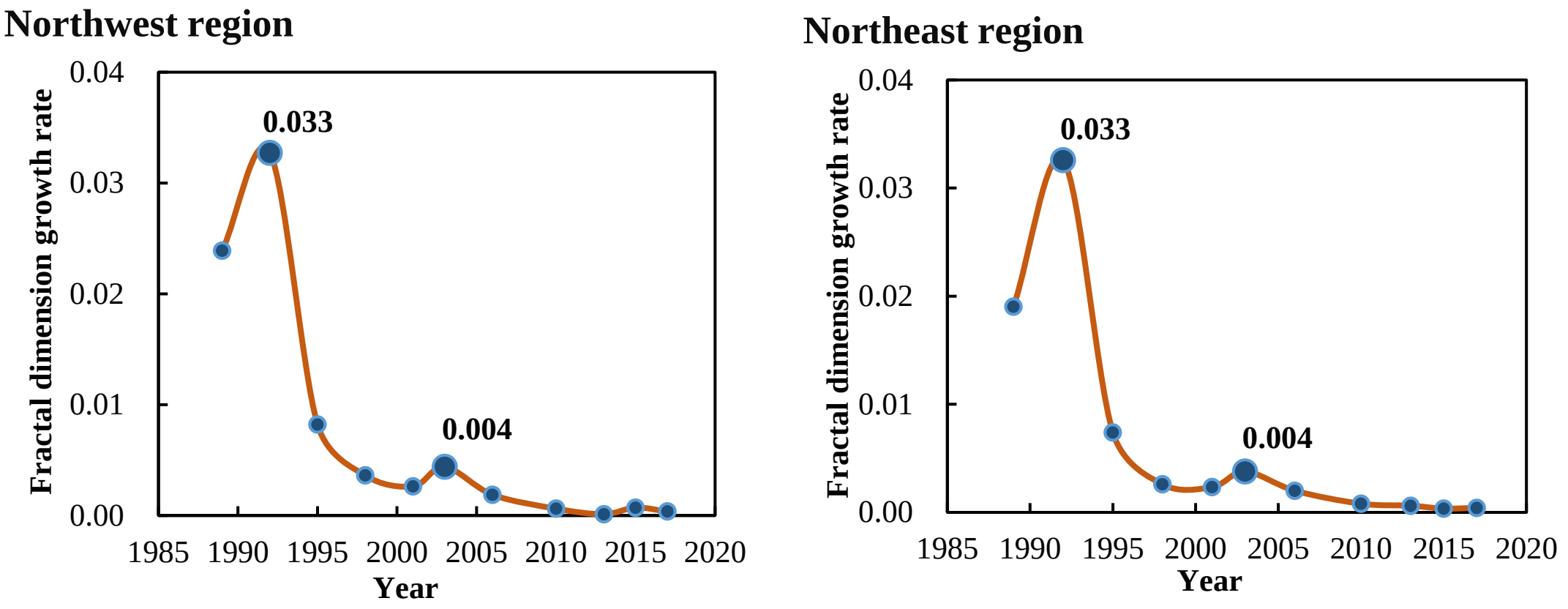

**Figure 2 The graph of the rate of growth of fractal dimension of four regions from 1986 to 2017**

## 3.2. Fractal dimension growth curves

The results of four regions fractal dimension sets to fit logistic model by ordinary least square (OLS) estimation and goodness-of-fit search (GOFS) is shown in **Fig.3**. It is obvious that fractal dimension growth curves of four regions in Shenzhen can be all very well fitted by first-order logistic function. The specific first-order logistic expressions and relevant parameters are shown in **Tab.2**, the goodness of fit $R^2$ of first-order logistic expressions for each region is very high. Meanwhile, we also can be simple to obtain the maximum capacity fractal dimension and predict the year of reaching maximum capacity by the logistic expression of each region. As shown in **Tab.2**, the maximum capacity fractal dimension of space of four study regions-Entire, center, Northwest and Northeast, in Shenzhen are 1.7905, 1.9, 1.86, and 1.8621, respectively, and the corresponding years of reaching above values are 2072, 2197, 2085, and 2082, respectively. it is simple to see that the center region in Shenzhen is both the largest maximum capacity fractal dimension of space and the longest time to reach year. But connecting with the practical situation, those values in fact are overvalued as within study areas, it includes the land that are completely inhospitable to man, such as river and high mountain.

**Tab.2.** Four regions logistics equation information table in Shenzhen, 1986-2017.

| **Region** | **Logistics equation** | $R^2$ | **Capacity** $D_{max}$ | $D_{max}$ **year** |
|---|---|---|---|---|
| Entire region | $\hat{D}(t)=\dfrac{1.7905}{1+0.5808*e^{-0.1129t}}$ | 0.9909 | 1.7905 | 2072 |
| Center region | $\hat{D}(t)=\dfrac{1.9000}{1+0.3110*e^{-0.0750t}}$ | 0.9868 | 1.9000 | 2097 |
| Northwest region | $\hat{D}(t)=\dfrac{1.8600}{1+0.4817*e^{-0.0992t}}$ | 0.9819 | 1.8600 | 2085 |
| Northeast region | $\hat{D}(t)=\dfrac{1.8621}{1+0.4581*e^{-0.1023t}}$ | 0.9801 | 1.8621 | 2082 |

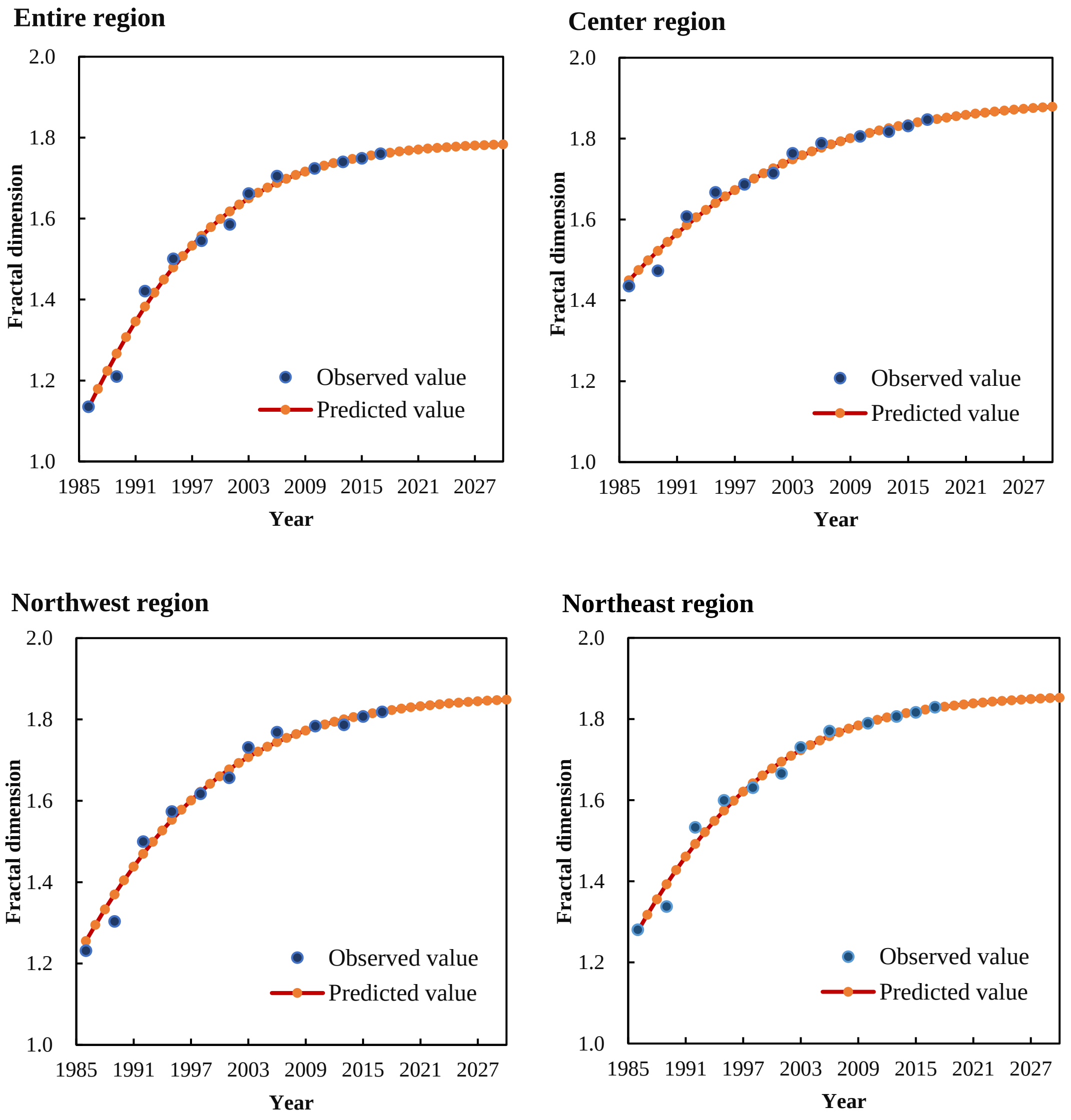

**Figure 3 The logistic patterns of fractal dimension growth of four study regions, 1986-2017**

## 3.3. Power law analysis

The results of the population size, GDP and fractal dimension of the entire Shenzhen from 1986 to 2017 are shown in **Fig.4** and **Fig.5**. Between population size, GDP and fractal dimension exist significant law relations, the values of goodness of fit $R^2$ are quite high, all above 0.96. The increase of fractal dimension with time usually indicates the urban sprawl. It usually relates to the factors of population size and the level of economic development (Rozenfeld *et al.*, 2008). It implies that both population size and GDP are the driving factors of promoting the urban sprawl in Shenzhen from 1986 to 2017. But the accelerating role of population to urban expansion is much greater than GDP in terms of their respective power exponent. It is also a problem worth thinking that whether or not both individual behavior and public policy play key roles in the Behind population and economic growth for a city? It will be discussed in the section 4 of this paper.

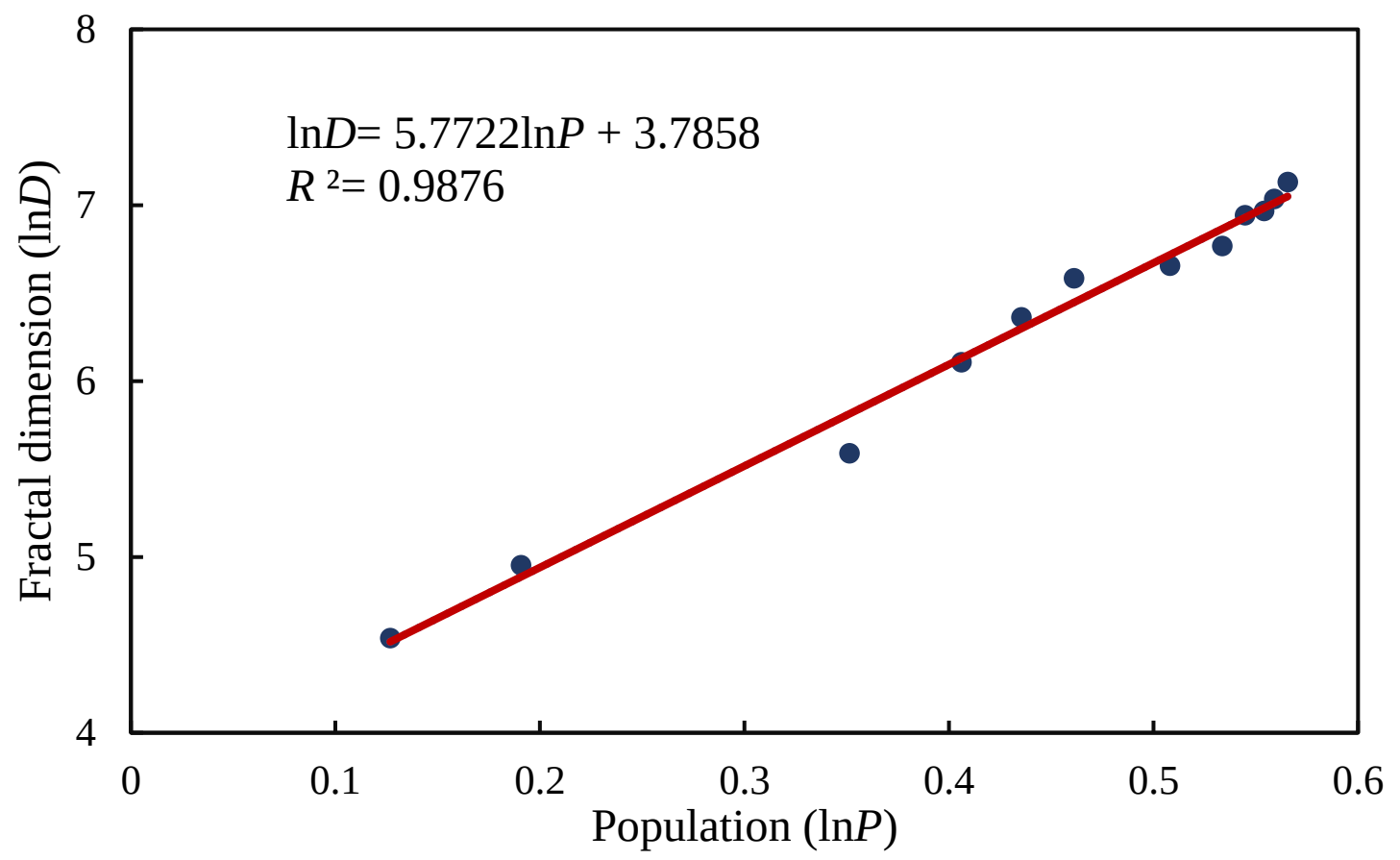

**Figure 4 Log-log plot between population (ln *P*) and fractal dimension (ln *D*) in Shenzhen, 1986-2017**

**Source:** Population data from Shenzhen Statistical Yearbook-2019

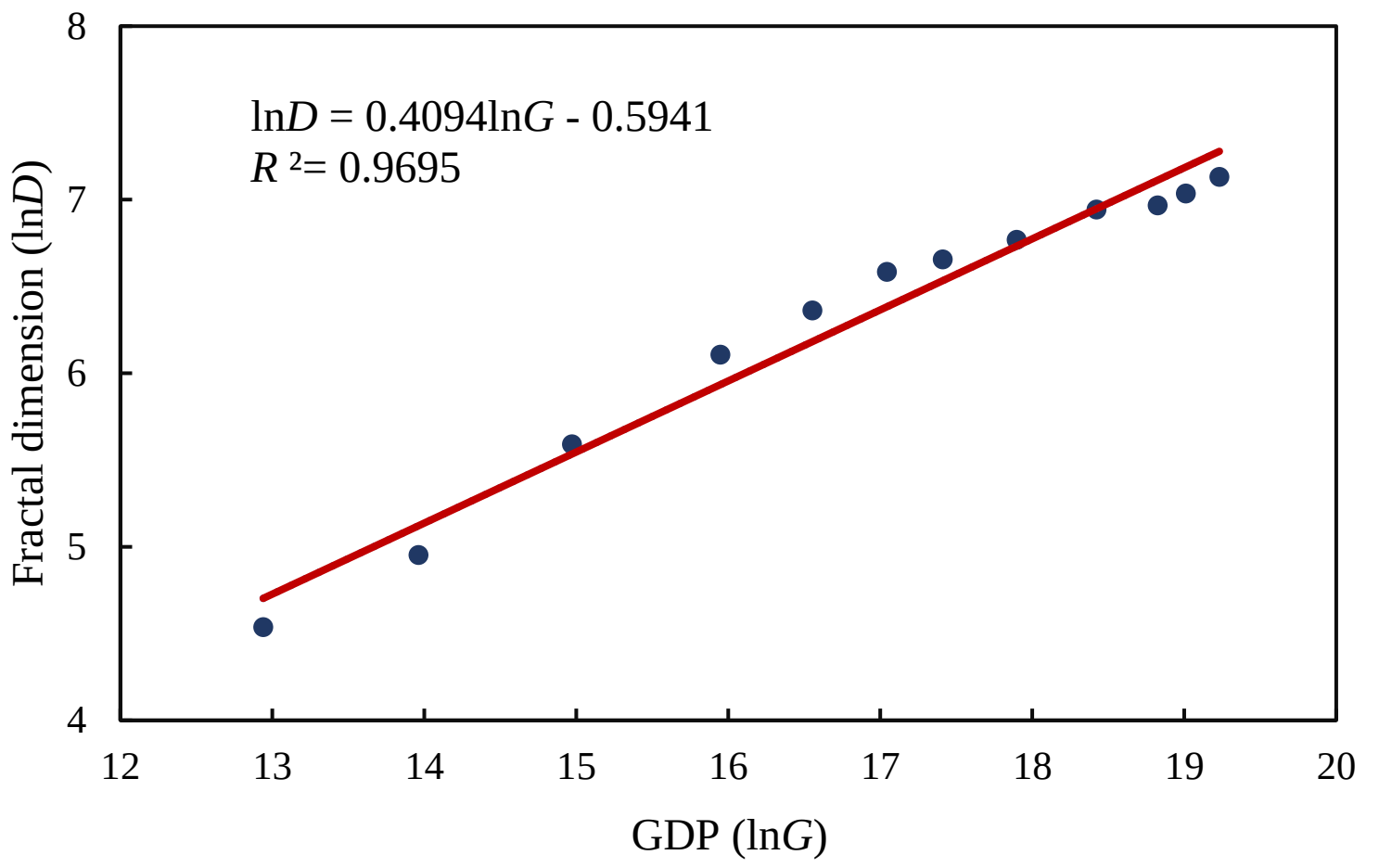

**Fig. 5. Log-log plot GDP (ln*G*) and fractal dimension (ln*D*) in Shenzhen, 1986-2017**

**Source:** Population data from Shenzhen Statistical Yearbook-2019

# 4. Discussion

By means of the calculation and analysis of fractal parameters, we can obtain a number of new knowledge about the city of Shenzhen. Some knowledge can be generalized to explain the spatio-temporal evolution of other cities. Where city fractals are concerned, Shenzhen differs from many cities in northern China (Chen and Huang, 2019). It is similar to an extent to the cities in Europe and American cities (Chen, 2012; Chen, 2018). This is revealing for us to understanding city development. The main points of above studies can be summarized as follows (Table 3). First, the urban form of Shenzhen possesses fractal structure. This suggests that the spatial order of this city have emerged by self-organized evolution. Second, different subareas of study takes on different fractal dimension values. This indicates spatial heterogeneity of Shenzhen's urban form, and spatial heterogeneity suggests multifractal scaling of city development. Third, fractal dimension values seems to descend from the center of city to the suburbs and exurbs. Especially at the early stages (1986-2001), the fractal dimension values of center region are significantly higher than fractal dimension values of northwest region and northeast region (Fig. 1, Tab. 1). This suggests a hidden circular structure of city development. The circular structure behind irregular urban form can be revealed by fractal dimension changes (White and Engelen, 1993). Fourth, the goodness of fit for fractal dimension estimation ascended over time. From 1986 to 2017, the *R* square value went up and up until it is close to 1 (Tab.1). This suggests that the fractal structure of Shenzhen became better and better gradually through self-organized evolution. Fifth, the fractal dimension growth curves of urban form can be modeled by conventional logistic function. This differs from the fractal dimension growth curves of the cities in northern China, but similar to those of the cities in western development countries. This maybe resulted from bottom-up urbanization process of southern cities in China dominated by self-organized evolution, which is associated with market mechanism. Sixth, the fractal dimension values approached the capacity parameters. All the fractal dimension in 2017 is close the maximum value, $D_{max}$. This suggests that the urban space of Shenzhen is filled to a great degree, and there is no many remaining space for future development.

**Table 3. The main results, findings of fractal studies and corresponding inferences or conclusions about Shenzhen city**

| **Results and findings** | **Inferences about Shenzhen** |
|---|---|
| Urban form bears fractal structure | Spatial order emerging from self-organization |
| Different parts bears different fractal dimension values | Possible multifractal structure |
| Fractal dimension values decay from center to edge | Hidden circular structure |
| Goodness of fit ascended over time | Fractal structure become optimized by self-organized evolution |
| Fractal dimension curves of urban growth meet logistic function | Bottom-up urbanization dynamics |
| Fractal dimension approached the maximum value | Space-filling of urban development is close to its limit |

Urban form and growth is associated with urbanization, the process of urbanization of a region seem to impact on the development of urban morphology. Urban form is one of important components of urbanization (Knox and Marston, 2009). The model of fractal dimension curve of urban growth is always consistent with the urbanization curve in a country or a region (Chen, 2018). Urbanization falls into two types: one is bottom-up urbanization, and the other, top-down urbanization (Zhou, 2010). The bottom-up urbanization is associated with market economy and chiefly dominated by the well-known "invisible hand" of free competition, while the top-down urbanization is associated with command economy and mainly dominated by the visible hand of administrative intervention (Chen, 2015). Different economic mechanisms and corresponding urbanization types have their own advantages and disadvantages. Bottom-up urbanization corresponds to self-organized evolution of cities. All cities can be treated as self-organized cities (Portuagli, 2000). However, self-organization processes of cities are influenced by political and economic system of a nation or a region. A fact is that China's southeast coastal areas opened earlier, and its economic development has been more strongly affected by the international community. This fact may account for the fractal dimension growth curves of Shenzhen's urban form.

The novelty of this work lies in two aspects. One is the investigation of different subareas. Shenzhen was divided into three overlapped subareas. Then we examine fractal structure and fractal dimension growth of the entire study area and three subareas. Although the similar way was once used by Benguigui *et al* (2000), the study area division of this paper bears its characteristics. The other is modeling fractal dimension curves of urban growth by conventional logistic function. This results in a new discovery that the fractal growth of southern cities differs from that of northern cities in China. This discovery leads to new understanding that the mode of urban growth corresponds to the mode of urbanization, and urbanization dynamics is dominated by the structure of economic system. The main shortcomings of this studies are as below. First, the data before 1986 is absent. We only found remote sensing images from 1986 and beyond. Thus we cannot identify the time in which the real peak of urban growth appeared. In fact, after 1986, the peak of the growth rate of urban land use in Shenzhen has passed. Second, the definition and division of study area are lack of sufficient objective bases. The principal criteria of study area and subareas are empirics and research objective. Third, only box-counting method was used. This method is suitable for measuring and estimating global fractal dimension. The local fractal dimension can be calculated cluster growing method, that is, by radius-area scaling (Frankhauser, 1998; White and Engelen, 1993). The cluster growing method can yield radial dimension (Frankhauser and Sadler, 1991). The radial dimension can be used to reflect urban growth from another angle of view.

Fractal dimension growth curves of urban form in four study regions of Shenzhen from 1986 to 2017 can be very well modeled with first-order logistic function (**Fig.3.**), which is the same with some western cities, such as London (UK), Tel Aviv (Israel) and Baltimore (USA) (Chen 2012, 2018). But Shenzhen city is different from the northern cities of China. The cities in the Beijing-Tianjin-Hebei region, such as Beijing, the fractal dimension growth curves of urban form all belong to the quadratic logistic curves. Moreover, the biggest difference between logistic curve and quadratic logistic curve is the rate of growth before the curve reaches the maximum capital value, logistic curve is much slower than quadratic logistic curve (**Fig.7.**). Nation's development stage and basic requirements and the public policy in a large extent is the main reason of influence the growth rate of curve. Conventional logistic Curve probably indicates market mechanism and bottom-up urbanization process, while quadratic logistic curve suggests government-led process and top-down urbanization. Obviously, Shenzhen is a city dominated by market economy and its self-organization feature is more prominent than those cities in northern China.

On the other hand, most cities in in northern China, often deeply impacted by planned economy for a long time. The planned economy in China, also known as the command economy, is generally referred to a kind of economic system in which the production, resource allocation and consumption are planned and decided by the government in advance. Especially in the early years of China's development, before the reformation and opening, land development can be regarded as a special product under the planned economy system, its development and utilization are all dominated by the government. Such urban development pattern can lead to the rapid expansion of Chinese urban form in a certain period of time, and the fractal dimension set of time series can be well fitted as the formal features of the quadratic logistic curve in **Fig.7.** Until the 1990s, China's began reform and opening up, begin to build a socialist market economy, set up some cities as special economic zones or the testing ground for developing the market economy. That way, the land development also gradually changed from the original government-lend mode to the mode of enterprise participation. Shenzhen, China is one of such representative cities. So Shenzhen is undoubtedly a special, notable city because it is not only a new, fast-growing city planned by government, but also an experimental city for government tests the operation of a market economy (Hao *et al.*, 2013). In spite of Shenzhen is a typical case with the characterizes of both the market economy and planned economy, according to **Tab.2**, it illustrates that market economy in Shenzhen has a bigger impact than public policy, or market economy dominates the development of Shenzhen rather than public policy. But, the role of public policy is irreplaceable in Shenzhen. It is perhaps precisely such development patterns that it quickly enabled its surprising economic development and population growth, became China's forefront of reform and opening up, led the development of Chinese economy.

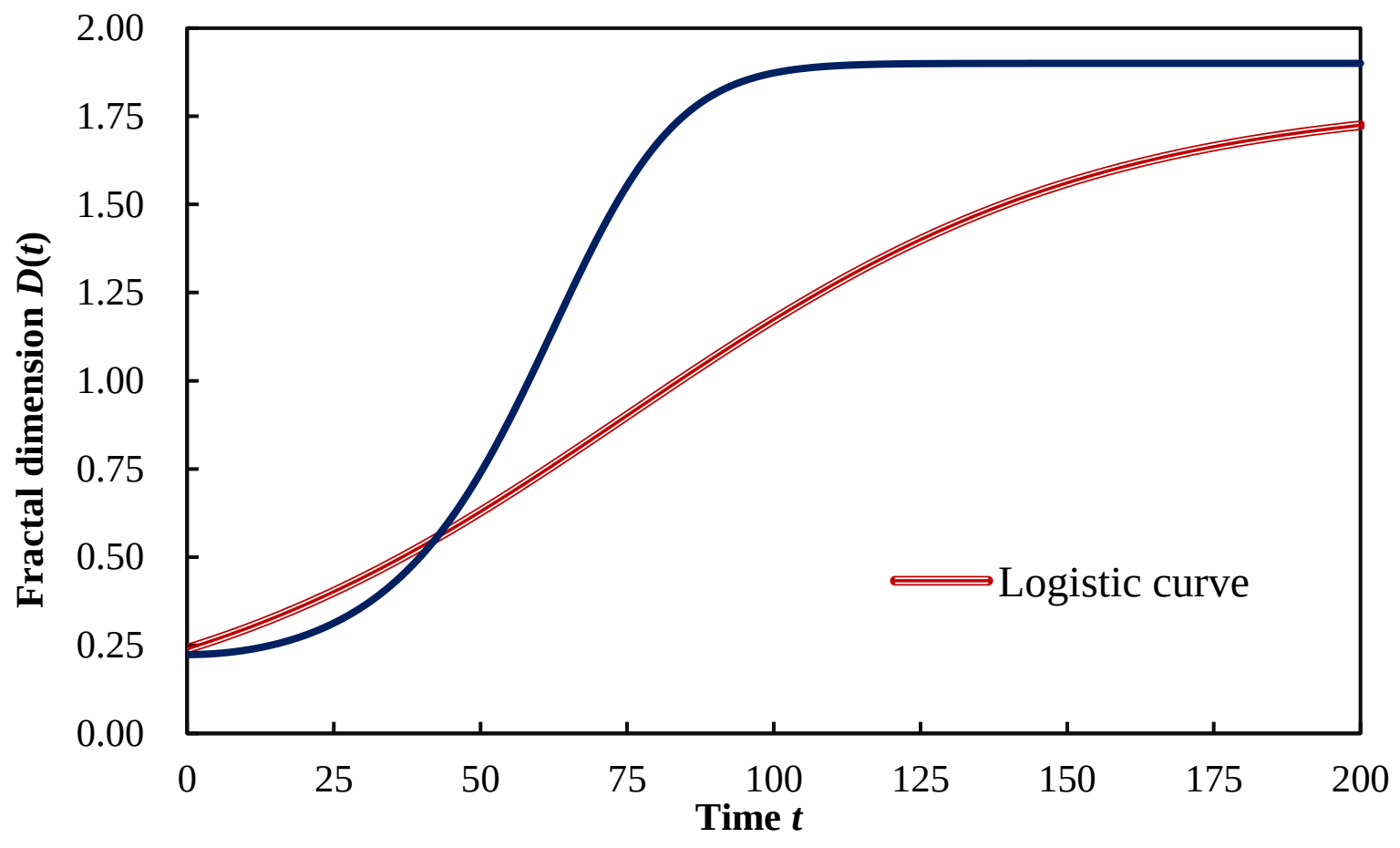

**Fig. 7.** Two types of fractal dimension growth curves of urban form (Chen and Huang, 2019).

**Note:** The logistic curve is based on London's model (Chen, 2018), while the quadratic logistic curve is based on Beijing's model.

In addition, urban form and growth presents the fractal structure at macro level not only shows the characteristics of self-similar and irregular of spatial distribution, but also reflects the process of self-organization at urban microcosmic level (Schelling, 1978; Allen, 1997; Portugali, 2000). It appears to endorse the view of Shen (2002) that individual parcels of urbanized areas can be geometrically planned and designed using Euclidean geometry, but planners or authorities can't regulate and predict human activities accurately. Meanwhile, the R square of goodness of fit of fractal model in Shenzhen from 1986 to 2017 increases gradually until it approached the limit of perfect fit value 1, and the corresponding standard error reflecting the fractal dimension error range decreases year by year. This supports the previous studies concluded that fractal is an evolutionary structure and optimizes gradually step by step through self-organization (Benguigui *et al*, 2000;

Chen and Wang, 1997; Chen and Feng, 2017); Fractal is the optimal structure of nature and can take up space most effectively (Chen, 2009). It may be a direction of urban planning in the future for using the thought and method of the fractal to optimize the space utilization of urban.

# 5. Conclusion

The scale-free spatial analysis of urban form revealed the fractal structure and evolution characteristics of Shenzhen city. This analysis is not only helpful for deep understanding the city of Shenzhen, but also is useful for understanding the regularities and dynamic mechanisms of urban evolution in different regions and even in different countries. The main conclusions of this study can be drawn as follows. **Firstly,** the urban form of Shenzhen city has the significant fractal feature, but the fractal structure has spatial heterogeneity. Different subareas of urban region show different fractal dimension values at the same time. Heterogeneity suggests multifractal scaling. Moreover, the fractal dimension decays from the center to the edge, indicating the circular structure of urban form. **Secondly,** the fractal dimension curves of urban growth of Shenzhen bear S-shape characteristics and can be modeled by conventional logistic function. This differs from the fractal dimension curves of the cities in northern China, which can be modeled with quadratic logistic function. Sigmoid curves suggest the squash effect of urban growth, but different logistic functions suggest different types of urbanization dynamics. The conventional logistic function indicates bottom-up urbanization, while the quadratic logistic function suggests top-down urbanization. **Thirdly,** the fractal structure of urban form of Shenzhen shows clear evolutionary process. Fractal dimension is a parameter of inference, which can be used to judge fractal structure. Meanwhile, fractal structure can be evaluated by the parameter of description, i.e., the goodness of fit. The $R$ square values of Shenzhen's fractal modeling went up and up over time until it approached 1. This suggests the fractal structure become optimized through self-organization. **Fourthly,** the fractal dimension of urban form in Shenzhen is approaching its limit, and the past urban development patters seem to be no longer sustainable. Fractal dimension values are tending towards the capacity value, which suggests that the space filling of Shenzhen is already near its limit, and there are few land resources available within the study area. It has to occupy the precious ecological resources or water resources if it continues to expand and extend. Thus, for Shenzhen city, new mode of development is needed in future.

## Acknowledgements

This research was sponsored by the National Natural Science Foundation of China (Grant No. 41671167). The support is gratefully acknowledged.

# References

Allen PM (1997). Cities and regions as evolutionary, complex systems. *Geographical Systems*, 4: 103-130

Arlinghaus SL (1985). Fractals Take a Central Place. *Geografiska Annaler: Series B, Human Geography*, 67: 83-88

Batty M (2008). The Size, Scale, and Shape of Cities. *Science*, 319: 769-771

Batty M (2009). Complexity and emergence in city systems: implications for urban planning. *Malaysian Journal of Environmental Management,* 10: 15-32

Batty M, Kim KS (1992). Form Follows Function: Reformulating Urban Population Density Functions. *Urban Studies,* 29: 1043-1069

Batty M, Longley M (1994). Fractal Cities- A Geometry of Form and Function. *Academic Press, London*

Batty M, Longley P, Fotheringham S (1989). Urban growth and form: scaling, fractal geometry, and diffusion- limited aggregation. *Environment and Planning A,* 21: 1447-1472

Batty M, Longley PA (1986). The fractal simulation of urban structure. *Environment and Planning A,* 18: 1143-1179

Batty M, Longley PA (1987a). Fractal-based description of urban form. *Environment and Planning B: Planning and Design,* 14: 123-134

Batty M, Longley PA (1987b). Urban Shapes as Fractals. *Area,* 19: 215-221

Batty M, Longley PA (1988). The morphology of urban land use. *Environment and Planning B: Planning and Design,* 15: 461-488

Batty M, Xie Y (1996). Preliminary evidence for a theory of the fractal city. *Environment and planning A,* 28: 1745-1762

Benguigui L (1998). Aggregation models for town growth. *Philosophical Magazine Part B* 77: 1269-1275

Benguigui L, Czamanski D, Marinov M (2001a). City growth as a leap-frogging process: an application to the Tel-Aviv Metropolis. *Urban Studies,* 38: 1819-1839

Benguigui L, Czamanski D, Marinov M (2001b). The dynamics of urban morphology: the case of Petah Tikvah. *Environment and Planning B: Planning and Design,* 28: 447-460

Benguigui L, Czamanski D, Marinov M, Portugali Y (2000). When and where is a city fractal? *Environment and Planning B: Planning and Design*, 27(4): 507–519

Benguigui LG, Czamanski D (2004). Simulation analysis of the fractality of cities. *Geographical Analysis,* 36: 69-84

Bettencourt LMA (2013). The origins of scaling in cities. *Science,* 340: 1438-1441

Boeing G (2018). Measuring the complexity of urban form and design. *Urban Design International (London, England)*, 23(4): 281-292

Chen YG (2009). A new model of urban population density indicating latent fractal structure. *International Journal of Urban Sustainable Development,* 1(1): 89-110

Chen YG (2012). Fractal dimension evolution and spatial replacement dynamics of urban growth. *Chaos, Solitons and Fractals,* 45: 115-124

Chen YG (2016). The evolution of Zipf's law indicative of city development. *Physica A*, 443: 555-567

Chen YG (2018). Logistic models of fractal dimension growth of urban morphology. *Fractals,* 26(3): 1850033

Chen, YG (2018). Logistic models of fractal dimension growth of urban morphology. *Fractals,* 26 (3).

Chen YG (2019). The solutions to uncertainty problem of urban fractal dimension calculation. *Entropy,* 21: 453

Chen YG (2020). Fractal modeling and fractal dimension description of urban morphology. *Entropy,* 22, 961

Chen YG (2020). Fractal Modeling and fractal dimension description of urban morphology. *Entropy* (Basel, Switzerland), 22(9): 961

Chen YG, Feng J (2017). A hierarchical allometric scaling analysis of Chinese cities: 1991-2014. *Discrete Dynamics in Nature and Society*, Volume 2017, Article ID 5243287

Chen YG, Huang LS (2019). Modeling growth curve of fractal dimension of urban form of Beijing. *Physica A: Statistical Mechanics and its Applications*, 523: 1038-1056

Chen YG, Wang JJ, Feng J (2017). Understanding the Fractal Dimensions of Urban Forms through Spatial Entropy. *Entropy*, 19: 600

Chen YG, Wang YJ (1997). A fractal study on interaction between towns in urban systems. *Bulletin of Science and Technology*, 13(4): 233–237 [In Chinese]

Dauphiné A (2013). *Fractal Geography*. Wiley-ISTE, Hoboken, NJ London

Encarnação S, Gaudiano M, Santos FC, Tenedório JA, Pacheco JM (2012). Fractal cartography of urban areas. *Scientific Reports*, 2:257

Fotheringham AS, Batty M, Longley PA (1989). Diffusion-limited aggregation and the fractal nature of urban growth. *Papers of the Regional Science Association*, 67: 55-69

Frankhauser P (1994). *La Fractalité des Structures Urbaines (The Fractal Aspects of Urban Structures)*. Economica, Paris

Frankhauser P (1998). The fractal approach: A new tool for the spatial analysis of urban agglomerations. *Population*, 10: 205-240

Frankhauser P (2004). Comparing the morphology of urban patterns in Europe. In: Borsdorf A, Zembri P (eds.). *European Cities - Insights on Outskirts: Structure*. COST, pp79-105

Frankhauser P, Sadler R (1991). Fractal analysis of agglomerations. In: Hilliges M (ed.). *Natural Structures: Principles, Strategies, and Models in Architecture and Nature*. Stuttgart: University of Stuttgart, pp 57-65

Hao BL (1986). Fractals and fractal dimensions. *Science*, 38 (1): 9-17 (in Chinese)

Hao P, Geertman S, Hooimeijer P, Sliuzas RV (2013). Spatial analyses of the urban village development process in Shenzhen, China. *International Journal of Urban and Regional Research*, 37: 2177-2197

Islam Z, Metternicht G (2003). Fractal dimension of multiscale and multisource remote sensing data for characterising spatial complexity of urban landscapes. *IEEE International Geoscience & Remote Sensing Symposium*, pp 1715-1717, DOI: 10.1109/IGARSS.2003.1294227

Jevric M, Romanovich M (2016). Fractal Dimensions of Urban Border as a Criterion for Space Management. *Procedia Engineering*, 165: 1478-1482

Keuschnigg M, Mutgan S, Hedström P (2019). Urban scaling and the regional divide. *Science Advances,* 5: 42

Knox PL, Marston SA (2009). *Places and Regions in Global Context: Human Geography (5th Edition)*. Upper Saddle River, NJ: Prentice Hall

Lagarias A, Prastacos P (2020). Comparing the urban form of South European cities using fractal dimensions. *Environment and Planning B, Urban Analytics and City Science*, 47(7): 1149-1166

Leyton-Pavez CE, Redondo JM, Tarquis-Alfonso AM, Gil-Martín JC, Tellez-Alvarez JD (2017). Fractal analysis of growing cities and its relationship with health centre distribution. *Proceedings of the Institute for System Programming of the RAS*, 29(2): 201-214

Li W, Wang Y, Peng J, Li GC (2005) Landscape spatial changes associated with rapid urbanization in Shenzhen, China. *International Journal of Sustainable Development and World Ecology,* 12: 314-325

Li ZX, Liu B, Wang R, Li Z (2013). Study on fractal characteristics of hilly city. *Journal of Applied Sciences (Asian Network for Scientific Information),* 13(7): 1155-1159

Liu SD, Liu SK (1994). *Solitary Wave and Turbulence*. Shanghai: Shanghai Scientific and Technological Education Publishing House (In Chinese)

Ma D, Guo RZ, Zheng Y, Zhao ZG, He FN, Zhu W (2020). Understanding Chinese urban form: the universal fractal pattern of street networks over 298 Cities. *ISPRS International Journal of Geo-Information*, 9(4): 192

Man W, Nie Q, Li ZM, Li H, Wu, XW (2019). Using fractals and multifractals to characterize the spatiotemporal pattern of impervious surfaces in a coastal city: Xiamen, China. *Physica A*, 520: 44-53

Mandelbrot BB (1967). How Long Is the Coast of Britain? Statistical Self-Similarity and Fractional Dimension. *Science,* 156: 636-638

Mandelbrot BB (1983). *The Fractal Geometry of Nature*. W.H. Freeman, New York

Ng MK (2003). Shenzhen. *Cities,* 20: 429-441

Ni C, Zhang S, Chen Z, Yan Y, Li Y (2017). Mapping the Spatial Distribution and Characteristics of Lineaments Using Fractal and Multifractal Models: A Case Study from Northeastern Yunnan Province, China. *Scientific Reports.* 7: 10511

Portugali J (2000). Self-Organization and the City. Berlin: Springer

Purevtseren M, Tsegmid B, Indra M, Sugar M (2018). The fractal geometry of urban land use: the case of Ulaanbaatar city, Mongolia. *Land* (Basel), 7(2): 67

Rastogi K, Jain GV (2018). Urban sprawl analysis using Shannon's entropy and fractal analysis: a case study on Tiruchirappalli city, India. *International Archives of the Photogrammetry, Remote Sensing and Spatial Information Sciences*, XLII-5: 761-766

Rozenfeld HD, Rybski D, Andrade JS, Batty M, Stanley HE, Makse HA (2008). Laws of Population Growth. *Proceedings of the National Academy of Sciences of the United States of America*, 105: 18702-18707

Salat S (2017). A systemic approach of urban resilience: power laws and urban growth patterns. *International Journal of Urban Sustainable Development*, 9: 107-135

Schelling TC (1978). *Micromotives and Macrobehavior*. Norton, London, New York

Shelberg MC, Moellering H, Lam N (1982). Measuring the fractal dimensions of empirical cartographic curves. *Auto Carto*, 5: 481-490

Shen GQ (2002). Fractal dimension and fractal growth of urbanized areas. *International Journal of Geographical Information Science,* 16: 419-437

Shenzhen Statistics Bureau (SSB) (2002). *Shenzhen Statistics Yearbook 2002*. Beijing: China Statistics Press

Shreevastava A, Rao PSC, McGrath GS (2019). Emergent self-similarity and scaling properties of fractal intra-urban heat islets for diverse global cities. *Physical Review E*, 100(3): 032142

Song Z, Yu L (2019). Multifractal features of spatial variation in construction land in Beijing (1985–2015). *Palgrave Communications*, 5(1): 1-15

Takayasu H (1990). *Fractals in the Physical Sciences*. Manchester: Manchester University Press

Thomas I, Frankhauser P, De Keersmaecker ML (2007). Fractal dimension versus density of built-up surfaces in the periphery of Brussels. *Papers in Regional Science*, 86: 287-308

Triantakonstantis DP (2012). Urban growth prediction modelling using fractals and theory of chaos. *Open Journal of Civil Engineering,* 2(2): 81-86

Tucek P, Janoska Z (2013). Fractal dimension as a descriptor of urban growth dynamics. *Neural Network World*, 23(2): 93-102

Versini PA, Gires A, Tchiguirinskaia I, Schertzer D (2020). Fractal analysis of green roof spatial implementation in European cities. *Urban Forestry and Urban Greening*, 49: 126629

Vicsek TA (1989). *Fractal Growth Phenomena.* World Scientific, Singapore

White R, Engelen G (1993). Cellular automata and fractal urban form: a cellular modelling approach to the evolution of urban land-use patterns. *Environment and Planning A,* 25: 1175-1199

Zhou YX (2010). *Exploration in Urban Geography*. Beijing: The Commercial Press [In Chinese]

# Appendices

The specific log-log plots of scaling relations of built-up area of four study regions in Shenzhen city between 1986 and 2017 are shown as in Appendixes A-D.

**Appendix A.** The log-log plots of scaling relations of built-up area in region 1.

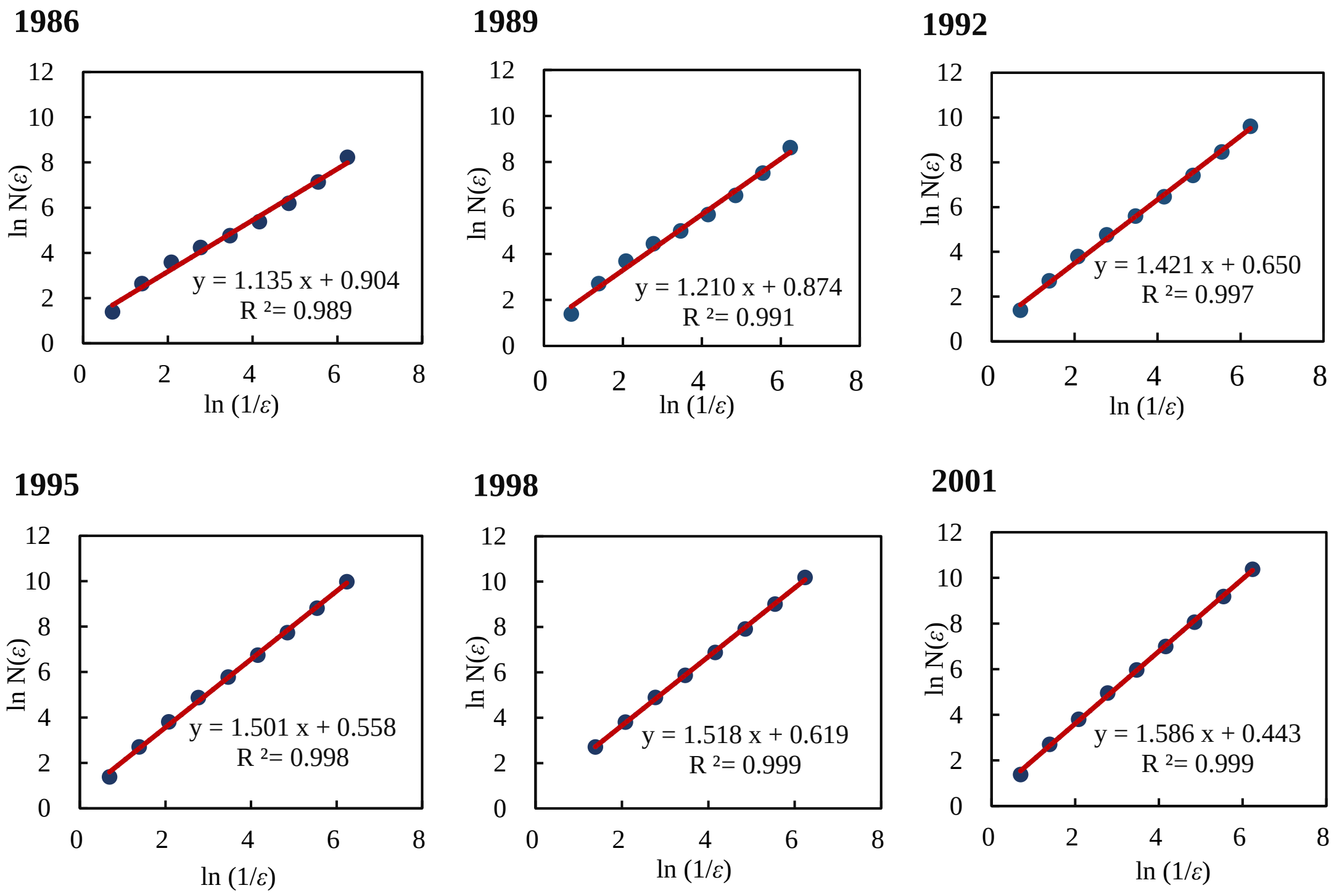

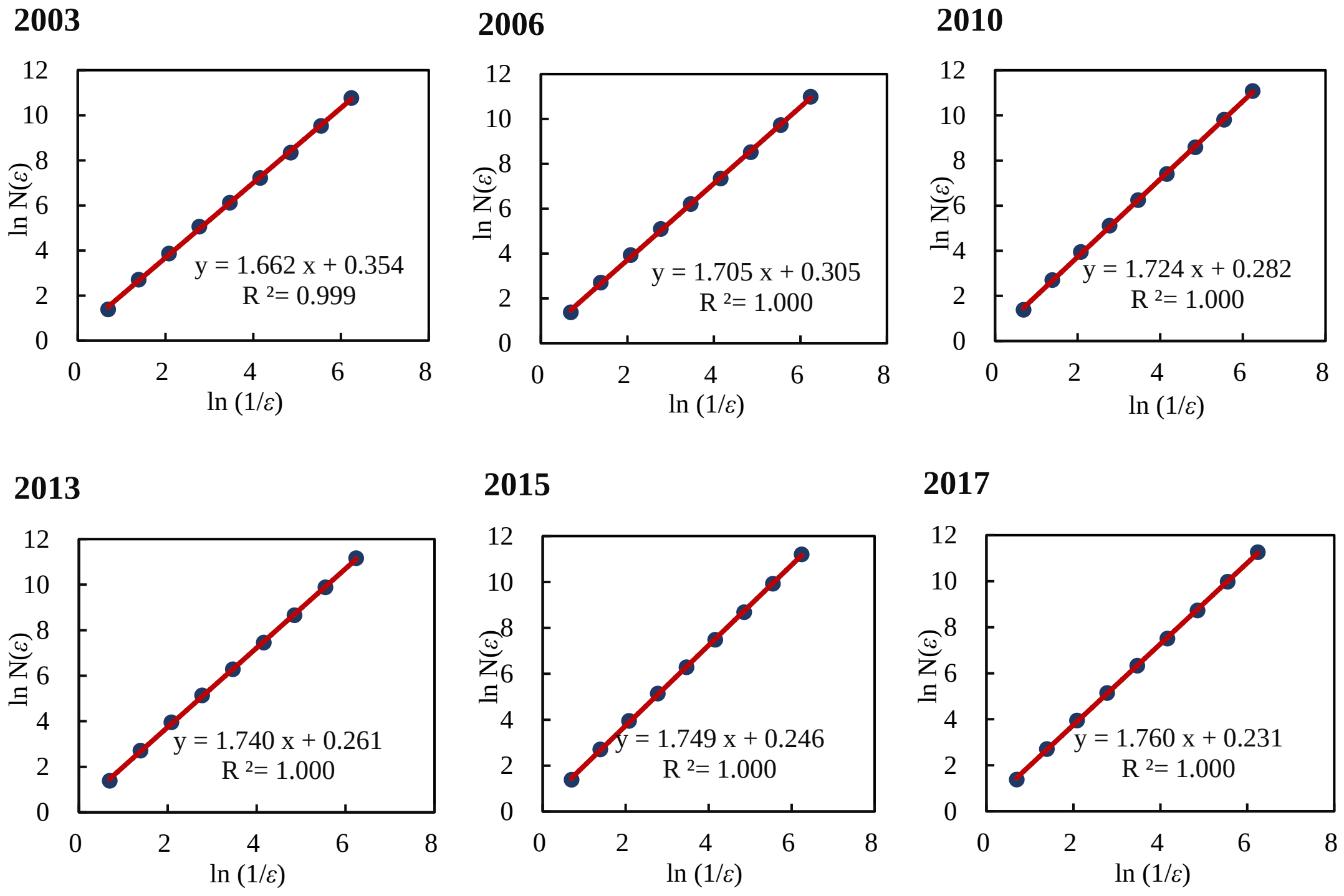

**Appendix B.** The log-log plots of scaling relations of built-up area in region 2.

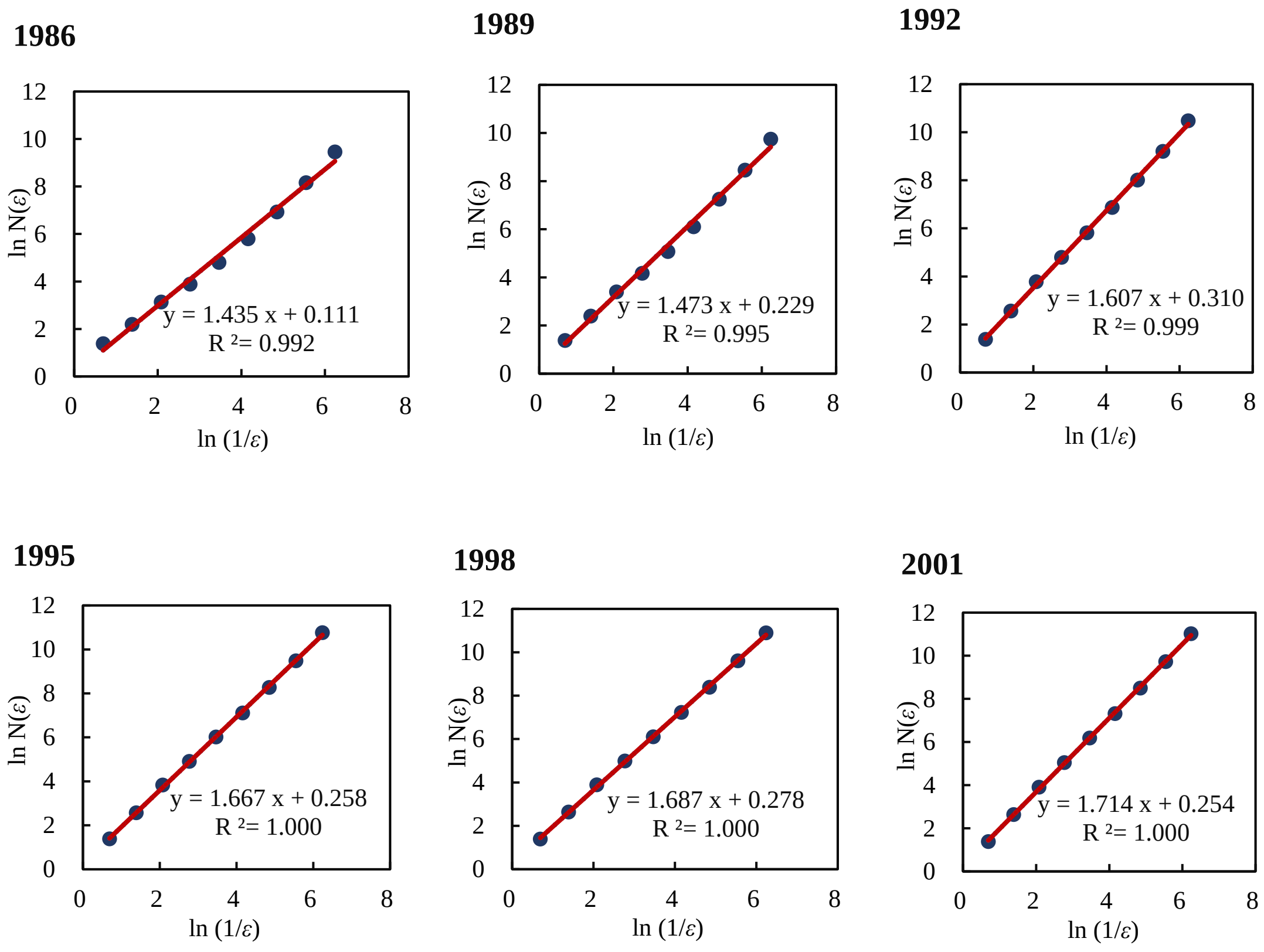

**2003**

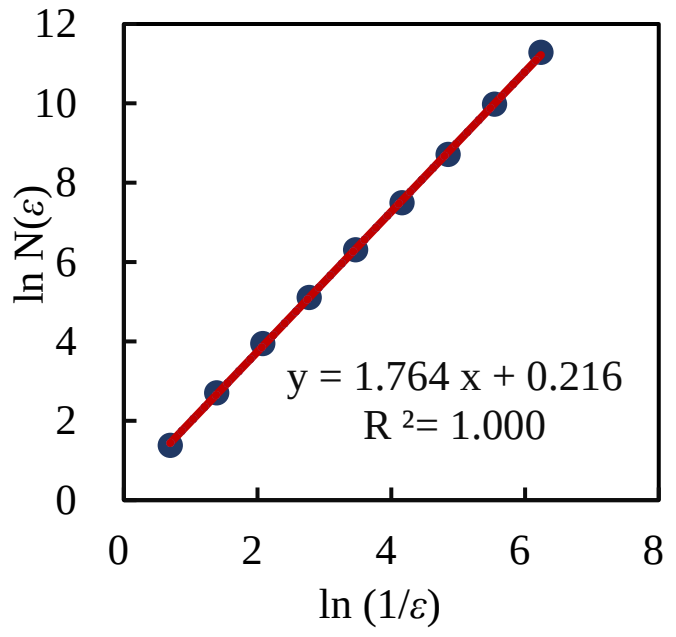

**2006**

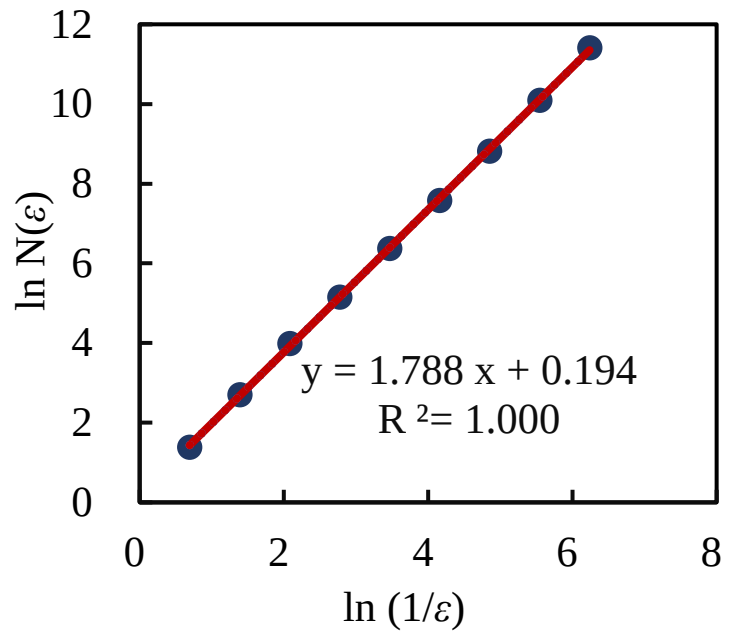

**2010**

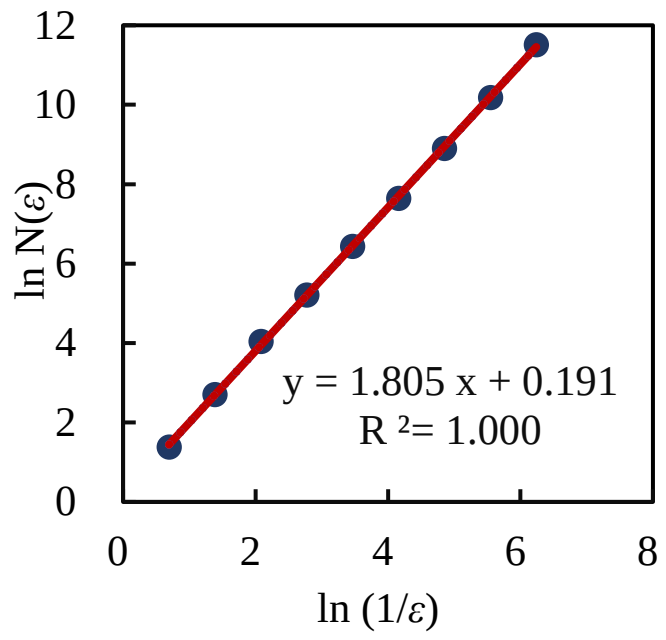

**2013**

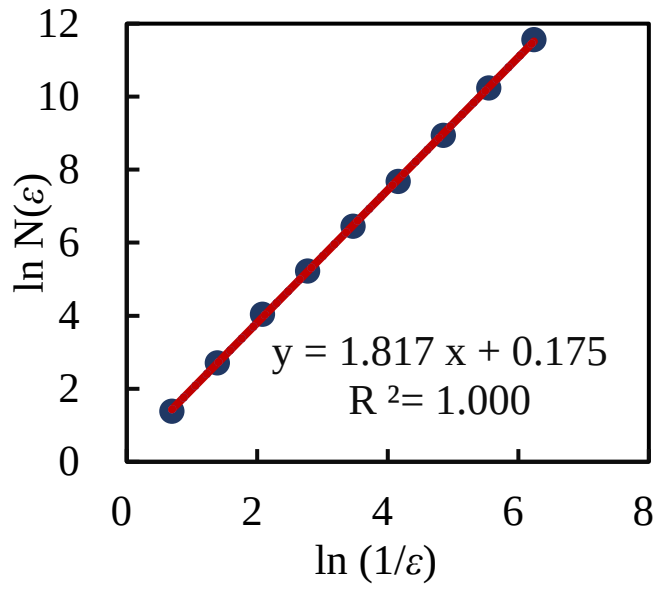

**2015**

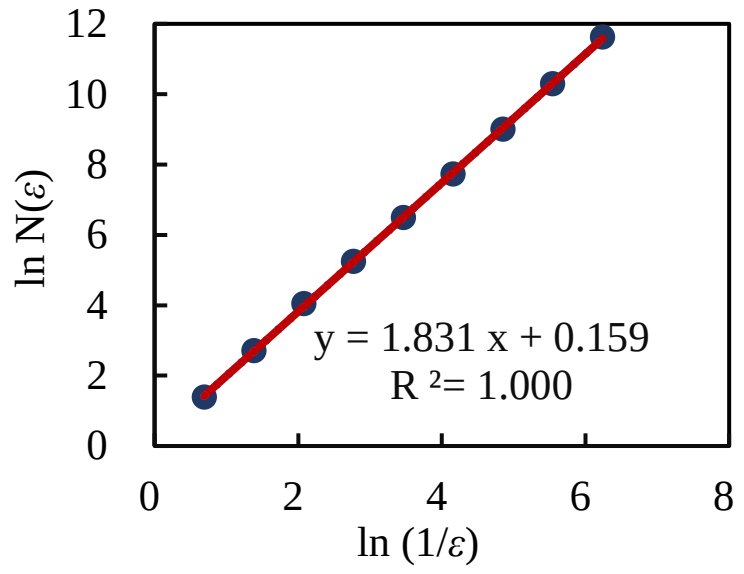

2017

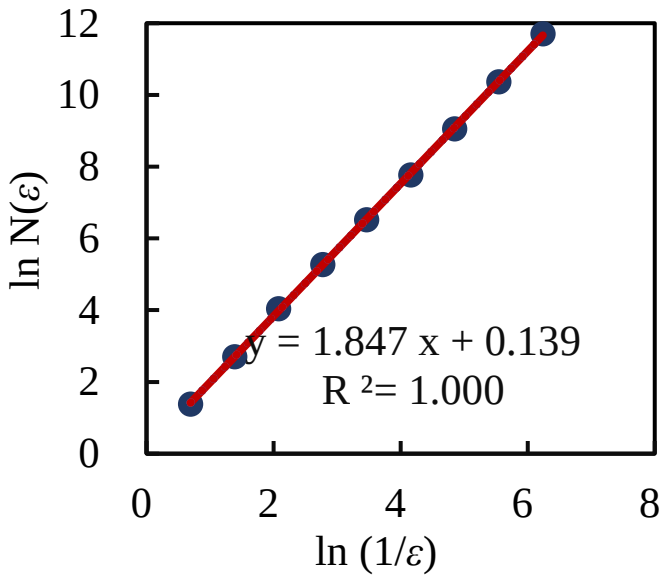

**Appendix C.** The log-log plots of scaling relations of built-up area in region 3.

**1986**

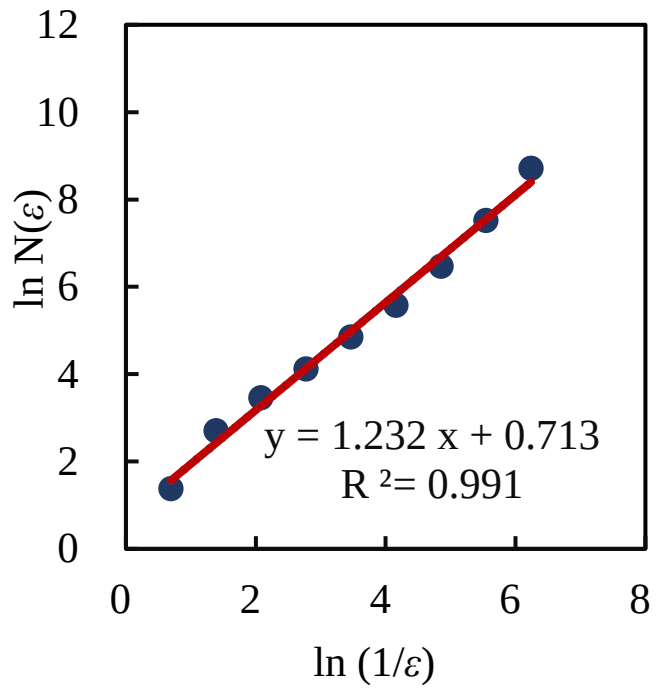

**1989**

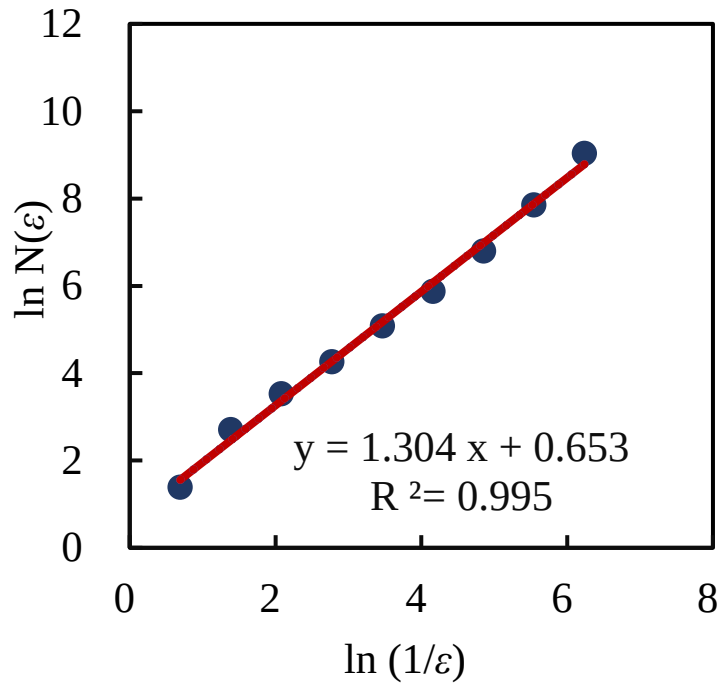

**1992**

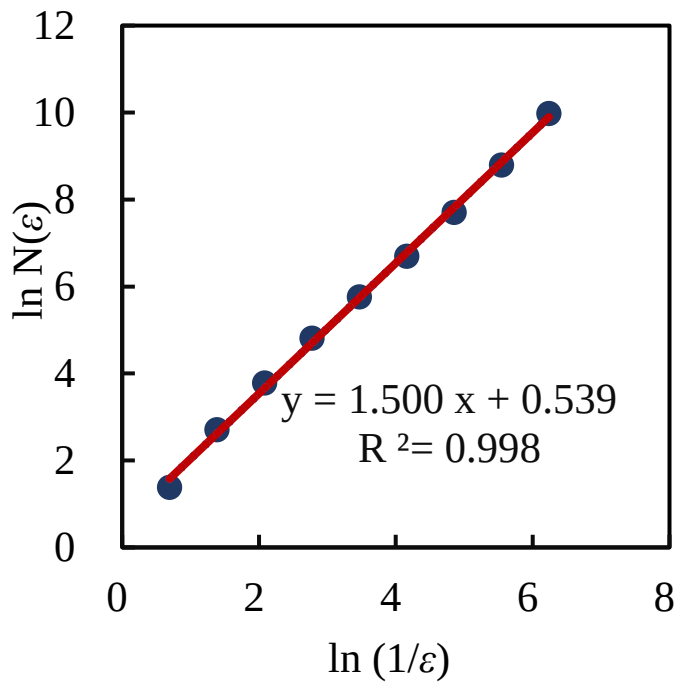

**1995**

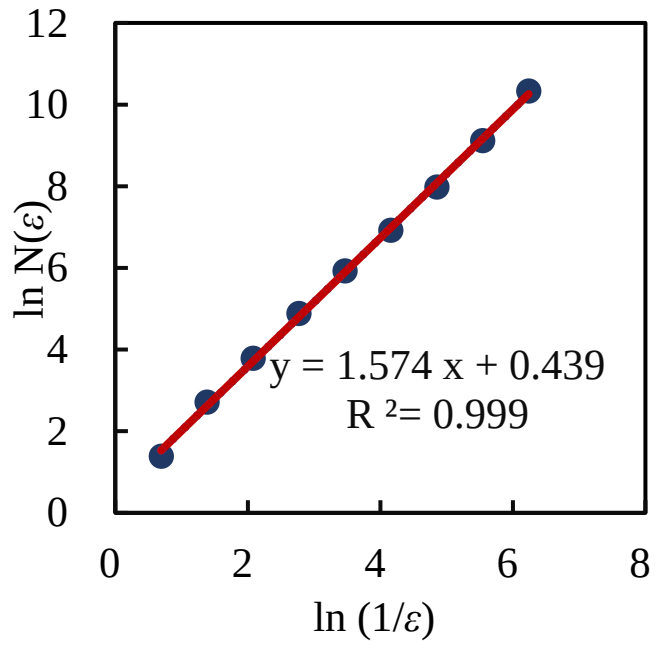

**1998**

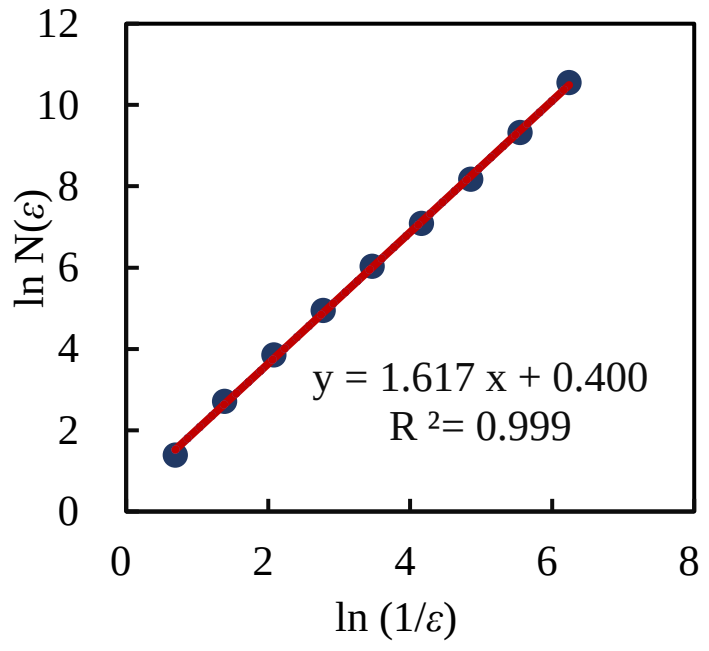

**2001**

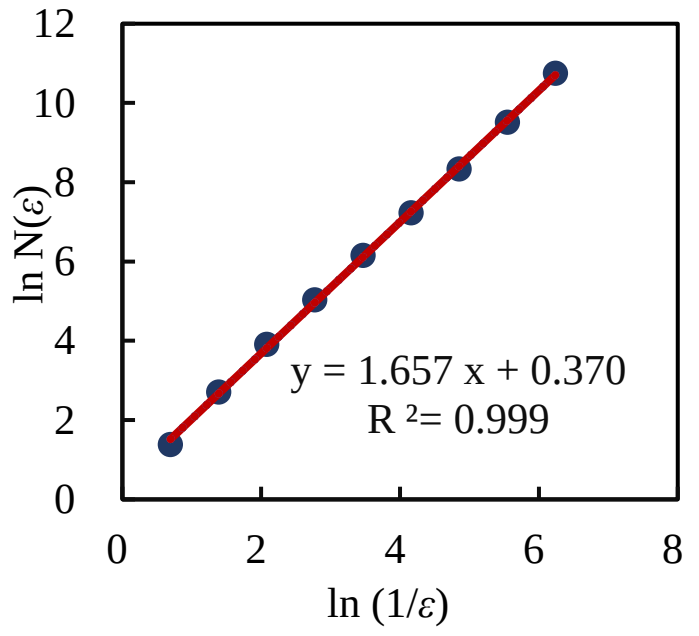

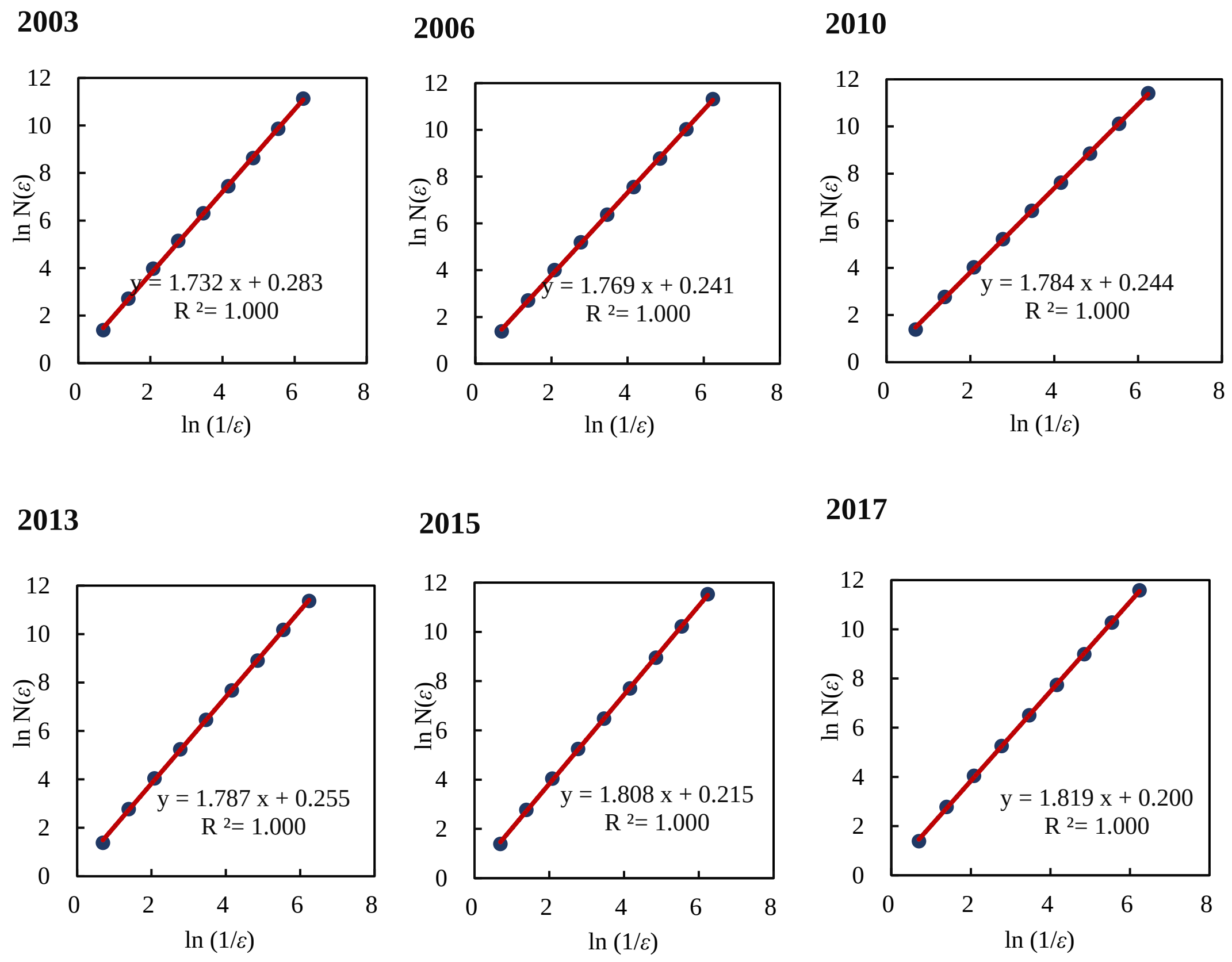

**Appendix D.** The log-log plots of scaling relations of built-up area in region 4.

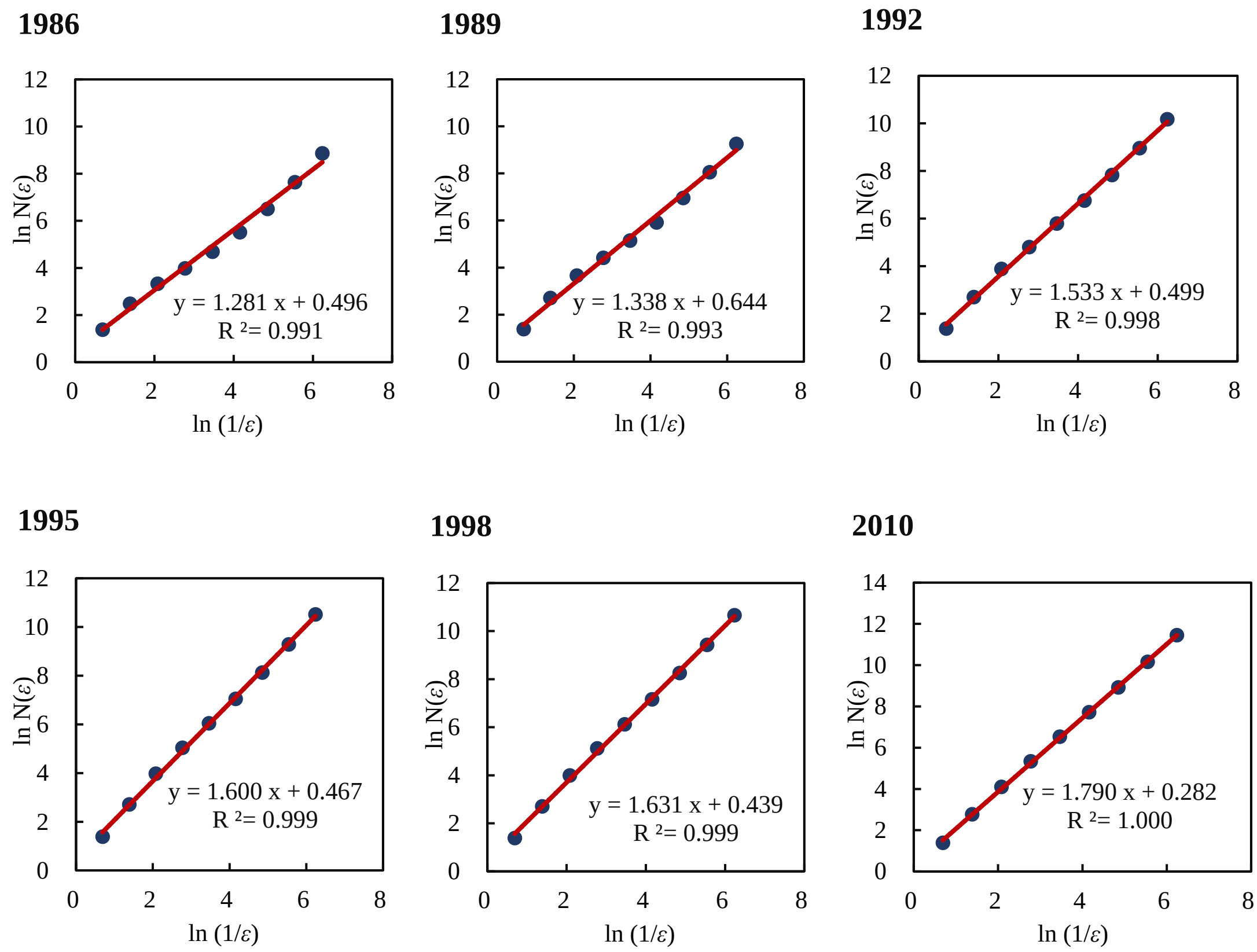

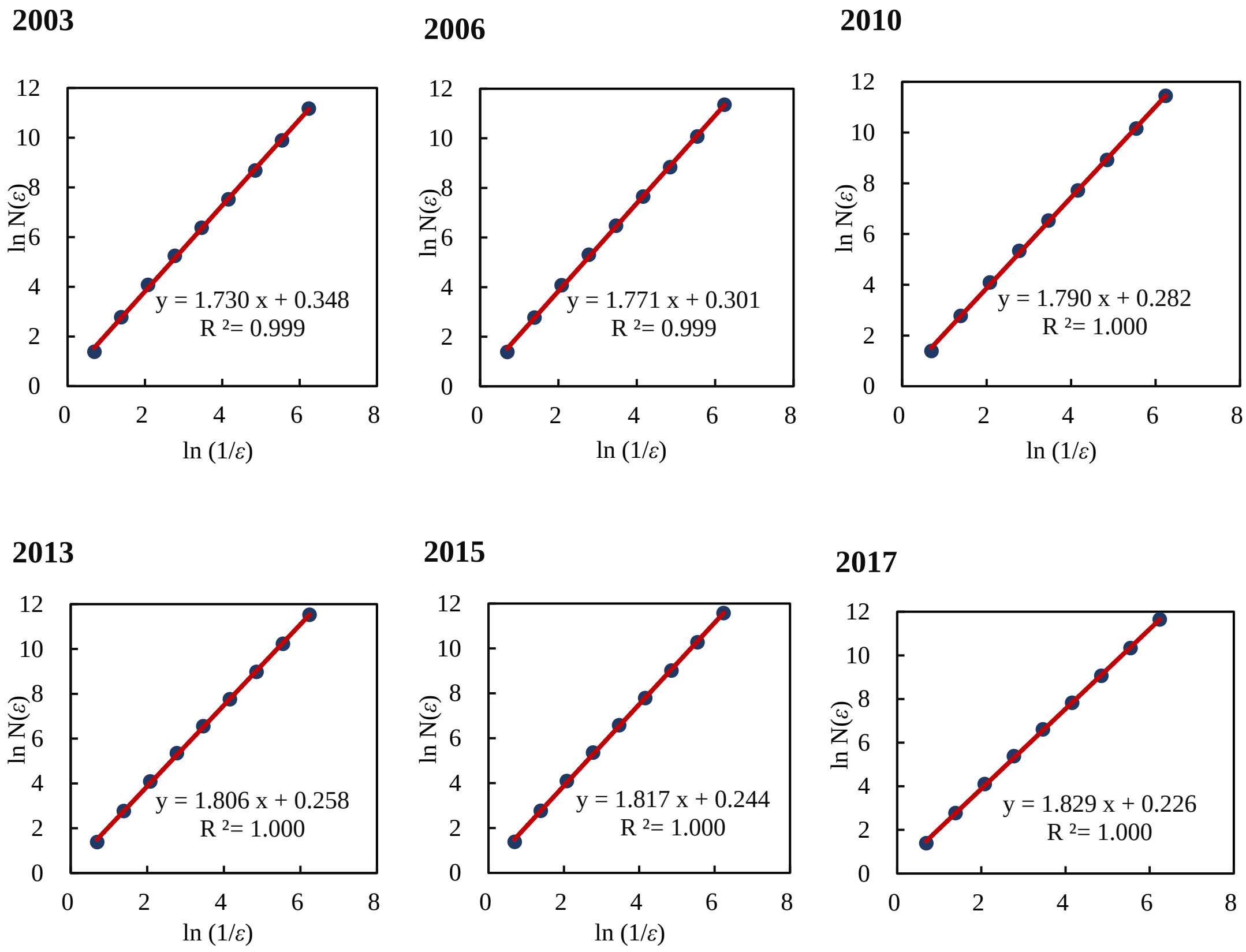
2003
y = 1.730 x + 0.348
R ²= 0.999
2006
y = 1.771 x + 0.301
R ²= 0.999
2010
y = 1.790 x + 0.282
R ²= 1.000
2013
y = 1.806 x + 0.258
R ²= 1.000
2015
y = 1.817 x + 0.244
R ²= 1.000
2017
y = 1.829 x + 0.226
R ²= 1.000
ln N(ε)
ln (1/ε)